\documentclass[nohyper,notoc]{JHEP}

\usepackage{amsmath,euscript,array,amssymb,cite} 
\newcommand{\startappendix}{
\setcounter{section}{0}
\renewcommand{\thesection}{\Alph{section}}}

\newcommand{\Appendix}[1]{
\refstepcounter{section}
\begin{flushleft}
{\large\bf Appendix \thesection: #1}
\end{flushleft}}

\def\aD{{\dot\alpha}}
\def\bD{{\dot\beta}}
\def\M{{{\cal M}}}
\def\N{{\cal N}}
\def\cN{{\cal N}}
\def\cO{{\cal O}}
\def\CM{{\cal M}}
\def\tr{{\rm tr}}
\def\trtwo{\tr^{}_2\,}
\def\Mbar{\bar{\cal M}}
\def\dalpha{{\dot\alpha}}

\def\wbar{\bar w}
\def\mubar{\bar\mu}
\def\abar{\bar a}
\def\sigmabar{\bar\sigma}
\def\etabar{\bar\eta}
\def\zetabar{\bar\zeta}
\def\mubar{\bar\mu}
\def\nubar{\bar\nu}
\def\Tr{{\rm Tr}}
\def\sst{\scriptscriptstyle}
\def\det{{\rm det}}
\def\SU{\text{SU}}
\newcommand{\BN}{\boldsymbol{N}}
\newcommand{\Bk}{\boldsymbol{k}}
\newcommand{\BL}{\boldsymbol{L}}

\def\Dbarslash{\,\,{\raise.15ex\hbox{/}\mkern-12mu {\bar\D}}}
\def\Dslash{\,\,{\raise.15ex\hbox{/}\mkern-12mu \D}}
\def\delslash{\,\,{\raise.15ex\hbox{/}\mkern-9mu \partial}}
\def\delbarslash{\,\,{\raise.15ex\hbox{/}\mkern-9mu {\bar\partial}}}

\def\Z{{\EuScript Z}}

\def\hf{{\textstyle{1\over2}}}

\def\l{\lambda} 
\def\s{\sigma}
 \def\betaR{\beta_{\sst R}}
 \def\betaIm{\beta_{\sst Im}}
\def\D{{\cal D}}
\def\Dbarslash{\,\,{\raise.15ex\hbox{/}\mkern-12mu {\bar\D}}}
\def\delslash{\,\,{\raise.15ex\hbox{/}\mkern-9mu \partial}}
\def\Dslash{\,\,{\raise.15ex\hbox{/}\mkern-12mu \D}}
\def\Skinst{S^k_{\rm inst}}
\def\dmuphys{d\mu^{k}_{\rm phys}}
\def\susic{supersymmetric}
\def\susy{supersymmetry}

\def\sqrtwo{\sqrt{2}\,}
\newcommand{\AdS}{AdS}
\newcommand{\U}{{U}}
\newcommand{\SO}{{SO}}

\def\bigR{{\rm I}\!{\rm R}}

\newcommand{\be}{\begin{equation}}
\newcommand{\ee}{\end{equation}}
\def\bea{\begin{eqnarray}}
\def\eea{\end{eqnarray}}
\newcommand{\EQ}[1]{\begin{equation} #1 \end{equation}}

\title{Instanton Calculations in the $\beta$-deformed AdS/CFT Correspondence}

\author{George Georgiou and Valentin V.~Khoze\\
Department of Physics and IPPP, University of Durham,
Durham, DH1 3LE, UK\\
E-mail: {\tt george.georgiou@durham.ac.uk}
{\tt valya.khoze@durham.ac.uk}}

\abstract{We consider non-perturbative effects in the $\beta$-deformed
$\cN=4$ supersymmetric gauge theory in the context of the AdS/CFT correspondence.
We concentrate on certain types of the $n$-point correlation functions of the Yang-Mills operators which
correspond to the lowest Kaluza-Klein modes propagating on the dual 
supergravity background found by Lunin and Maldacena in [1].
In particular, 
we calculate all multi-instanton contributions to these correlators
in the  $\beta$-deformed SYM and find a compelling agreement with the 
results expected in supergravity.}

\preprint{{\tt hep-th/0602141}\\IPPP/06/18\\
DCPT/06/36}

\begin{document}

\section{Introduction} 

$\beta$-deformations of the $\cN=4$ supersymmetric Yang-Mills define a
family of conformally-invariant four-dimensional $\cN=1$ supersymmetric gauge theories.
Remarkably, these $\beta$-deformed theories mirror many non-trivial characteristic features
of the $\cN=4$ SYM, including the S-duality, the AdS/CFT correspondence and
the perturbative large-N equivalence to the parent $\cN=4$ theory, thus providing a continuous class
of interesting generalizations of the $\cN=4$ SYM. One may expect that by studying
properties of the $\beta$-deformed theories and, in particular, the dependence
of observables on the continuous deformation parameter $\beta$ we can
understand better the gauge dynamics of this class of theories and of the $\cN=4$ as well.

Dualities between gauge and string theories have been studied intensively for more than three decades. 
The AdS/CFT correspondence formulated in \cite{Maldacena,GKP,Witten150} provides a concrete realization of such a duality.
In its original formulation, the AdS/CFT duality relates the
string theory on a curved background $AdS_5 \times S^5$ to the ${\cal{N}}=4$ supersymmetric 
Yang-Mills theory living on the boundary of $AdS_5$. Understanding this duality in detail 
beyond the BPS and the near BPS limits remains a challenge mainly due to the fact that one 
has to deal with the weak-to-strong coupling correspondence.

The AdS/CFT duality extends to the $\beta$-deformed theories \cite{LM} where
it relates the $\beta$-deformed $\cN=4$ SYM and the supergravity on the deformed
$\AdS_5\times \tilde{S}^5$ background.
The gravity dual was found by Lunin and Maldacena in Ref.~\cite{LM}, and this provides
a precise formulation of the AdS/CFT duality in the deformed case,
which can be
probed and studied using instanton methods developed earlier for the 
$\cN=4$ case in \cite{BGKR,DKMV,DHKMV,MO3} and in \cite{BG,Green,Review,GK}.

Several perturbative calculations in the $\beta$-deformed theories were carried out
recently in \cite{FG,PSZ,MPSZ,VVK} where it was noted that there are many similarities between
the deformed and the undeformed theories which emerge in the large number of colours limit.
In particular, in \cite{VVK} it was shown that in perturbation theory
there is a close relation between the scattering amplitudes in the $\betaR$-deformed and
in the original $\cN=4$ theory. This correspondence holds in the
large-$N$ limit and to all orders in (planar) perturbation theory. It states that
for real values of $\beta$
all amplitudes in the $\beta$-deformed theory are given
by the corresponding $\cN=4$
amplitudes times an overall $\beta$-dependent phase factor. The phase factor depends only on the
external legs and is easily determined for
each class of amplitudes \cite{VVK}. 
It follows from these considerations \cite{VVK} that the
recent 
proposal of Bern, Dixon and Smirnov \cite{BDS} which determines all multi-loop MHV planar
amplitudes in the $\cN=4$ superconformal Yang-Mills theory
can be carried over to a wider family of gauge theories obtained by real $\beta$-deformations
of the $\cN=4$ Yang-Mills.

The purpose of this paper is to consider non-perturbative instanton effects
in the $\beta$-deformed theories and in the context of the AdS/CFT correspondence.
The Lunin-Maldacena example \cite{LM} of the AdS/CFT duality
relates the large-$N$ limit of the $\beta$-deformed $\cN=4$ gauge theory to the
type IIB string theory on the appropriately $\beta$-deformed $\AdS_5\times \tilde{S}^5$ background.
The $\beta$-deformed gauge theory is living on the 4-dimensional boundary of the
$\AdS_5$ space. It is expected that each chiral primary operator (and its superconformal descendants)
in this boundary conformal theory
corresponds in supergravity to a particular Kaluza-Klein mode on the deformed sphere $\tilde{S}^5$.
In this paper we will consider only the operators which correspond to the supergravity states which
do not depend on $\tilde{S}^5$ coordinates, i.e. which are
the lowest Kaluza-Klein modes on the deformed sphere. Furthermore, to simply the derivations,
we will restrict ourselves to a particular class of such operators, considered previously
in \cite{BGKR,MO3}, and to the minimal in $n$ classes of their correlators $G_n$.  

Following the approach developed in Ref.~\cite{MO3},
we will evaluate all multi-instanton contributions to these correlation functions
in the appropriate large-$N$ scaling limit\footnote{The large-$N$ limit appropriate for the comparison
with the supergravity solution of \cite{LM}, will also require that the deformation parameter $\beta$ 
is kept small. Hence, starting from Section {\bf 5} we will take $N\to \infty$ and $\beta \to 0.$} 
and to the leading non-vanishing order in perturbation theory around
instantons. 
We will show that
these correlation functions in the $\beta$-deformed $\cN=4$ SYM 
are in precise correspondence with the supergravity expectations.
More precisely, we will be able to reproduce a class of leading higher-derivative
corrections to the supergravity effective action, $S_{\rm eff},$
from Yang-Mills instantons.
In particular, we will see that the multi-instanton 
contributions to $G_n$ will reconstruct
the appropriate moduli forms $f_n(\tau, \bar\tau)$ present in the $S_{\rm eff}.$
This is a particularly non-trivial observation since the dilaton-axion $\tau$ 
parameter in the {\it deformed} supergravity solution is not anymore equal to the complexified
coupling $\tau_0$ of the gauge theory. The dilaton $\phi$ is, in fact, a 
non-trivial function of the coordinates
$\mu_i$ on the deformed sphere \cite{LM}
\be
e^{\phi} = G^{1/2}(\beta; \mu_1,\mu_2, \mu_3) \,\cdot\, {g^2 \over 4 \pi}
\label{E11}
\ee
Here $g^2$ is the Yang-Mills coupling constant and $G$ is the function appearing in the 
Lunin-Maldacena solution in Eqs.~\eqref{GammaDeform}-\eqref{GLMdef} below.
It will turn out, that in the $\beta$-deformed $\cN=4$ SYM,
a proper inclusion of instanton collective coordinates will effectively upgrade
the usual exponent of the $k$-instanton action $e^{2\pi i k\tau_0}$
into the required expression $e^{2\pi i k\tau}.$
 
The rest of our findings parallels those in Ref.~\cite{MO3}.
We shall find that in the appropriately taken large-$N$ scaling limit,
the $k$-instanton collective coordinate measure 
has a geometry of a single
copy of the 10-dimensional space $\AdS_5\times \tilde{S}^5.$
We shall also observe that this $k$-instanton measure includes the partition function
of the $SU(k)$ matrix model, thus matching the description of the D-instantons 
as D(-1) branes in
string theory.
In the Appendix we show that 
the full Yang-Mills $k$-instanton integration measure in the deformed theory
is equivalent to the partition function of $k$ D-instantons in the corresponding
string theory where $\beta$-deformations are introduced via star products.

In Sections {\bf 2-7} we consider the transformations with a real
deformation parameter $\beta = \betaR.$ The generalization to the case
of complex deformations is carried out in Section {\bf \ref{sec:gencompl}}.
That Section contains the main results of the paper.
There we  explain how our instanton approach and the matching between the supergravity
and the gauge theory expressions found at real values of $\beta$
also apply for complex deformations. 

The real-$\beta$-deformation of the $\cN=4$ supersymmetric gauge theory
is described by the superpotential
\be
\label{superpot2}
i g \,\Tr( e^{ i \pi \betaR } \Phi_1 \Phi_2 \Phi_3 - e^{-i \pi \betaR } \Phi_1 \Phi_3 \Phi_2 )\ ,
\ee
where $\Phi_i$ are chiral $\N=1$ superfields.
The resulting superpotential preserves the $\N=1$ supersymmetry of the original $\N=4$ SYM
and leads to a theory with a global $U(1)\times U(1)$ symmetry 
(in addition to the usual $U(1)_R$ R-symmetry of the $\cN=1$ susy)
\cite{LM}
\bea \nonumber
U(1)_1:& ~~~~&(\Phi_1,\Phi_2 ,\Phi_3 ) \to (\Phi_1,e^{i\varphi_1}\Phi_2 ,e^{-i \varphi_1} \Phi_3 )
\\ \label{u1s}
U(1)_2:&~~~~&(\Phi_1,\Phi_2 ,\Phi_3 ) \to (e^{-i \varphi_2} \Phi_1,e^{i\varphi_2}\Phi_2 , \Phi_3 )
\eea
It is known \cite{MPSZ,VVK}
that \eqref{superpot2} describes an exactly marginal deformation
of the theory in the limit of large number of colors. 

The more general case of complex $\beta$-deformations will be discussed in detail in 
Section {\bf \ref{sec:gencompl}}. These deformations are characterized by the superpotential \eqref{superpotC}.
The resulting deformed gauge theory is an $\cN=1$ supersymmetric gauge theory with a global symmetry
\eqref{u1s}. Importantly, the complex $\beta$-deformation is exactly marginal if the parameters of the deformation
satisfy a certain condition -- the Leigh-Strassler constraint \cite{LS}.  For the real $\betaR$ case
this constraint is trivial 
in the large $N$ limit.
Another important feature of the $\beta$-deformed gauge theory is the $SL(2,Z)_s$ duality.
It was pointed in \cite{DHK} that under the $SL(2,Z)_s$ transformations, the coupling constant and
the deformation parameter $\beta$ must transform as modular forms.
Other aspects of the $\beta$-deformed gauge theory have been studied in
in Refs.~\cite{Berenstein:2000hy,Berenstein:2000ux,Dorey:2003pp,Hollowood:2004ek,Dorey:2004xm,
Frolov:2005ty,Frolov,Beisert:2005if,Frolov:2005iq,Rossi:2005mr,Kuzenko:2005gy,Hernandez:2005xd,Chen:2005sb,Chen:2006bh}.

\section{Supergravity Dual} \label{sec:sugra}

The gravity dual of the $\beta$-deformed 
${\cal{N}}=4$ gauge theory was identified by Lunin and Maldacena in \cite{LM}.
The $U(1) \times U(1) $ global symmetry \eqref{u1s} of the $\beta$-deformed 
SYM plays an important r{\^o}le in this approach. One starts with the original 
$AdS_5 \times S^5$ background and compactifies it on a two-torus in such a way that the
isometries of the torus  match with 
the global $U(1) \times U(1) $ symmetry in gauge theory. The idea \cite{LM} is then to 
use the $SL(2,R)$ symmetry of type IIB supergravity compactified along the two-torus 
to generate a new solution of the supergravity equations.
The $SL(2,R)$ transformation acts on the 
complex parameter $\tau=B_{12}+i\sqrt{g}$ of the original gravitational theory.
Here $ B_{12}$ is the NS-NS two-form field along the torus directions, 
and $g$ is the determinant of the metric on the torus.
The  $SL(2,R)$ acts on this torus $\tau$-parameter as follows
\bea
\tau \longrightarrow  \frac{\tau}{1+\betaR \tau}
\eea 
The geometry obtained in 
this way is (in the string frame) the product of $AdS_5 \times \tilde{S}^5$, 
where $\tilde{S}^5$ is a deformed five-sphere. In what follows we write down 
only the part of the supergravity solution which will be relevant for our purposes,\footnote{As
mentioned earlier we also restrict to real values of $\beta = \betaR$ which was called
$\gamma$ in \cite{LM}. We will comment on the complex deformations in Section {\bf \ref{sec:gencompl}}.}
\bea\label{GammaDeform}
ds^2_{str} &=& R^2 \left[ ds^2_{AdS_5} +  \sum_i ( d\mu_i^2  + G \mu_i^2 d\phi_i^2) 
+ \hat \gamma^2G \mu_1^2\mu_2^2\mu_3^2 (\sum_i d\phi_i)^2 \right] \ ,
\\ \nonumber
\\
e^{\phi} &=& e^{\phi_0} G^{1/2} 
\label{dildef}\ ,\\ \nonumber
\\
\label{GLMdef}
G^{-1} &=& 1 + \hat \gamma^2 (\mu_1^2 \mu_2^2 + \mu_2^2 \mu_3^2 + \mu_1^2 \mu_3^2)\ , \qquad
\hat \gamma := R^2 \betaR \  ,\qquad R^4 = 4 \pi e^{\phi_0} N
\eea
The deformed five-sphere in the supergravity solution above
is parameterized by the three radial variables $\mu_i$, which satisfy the condition $\sum_{i=1}^3\mu_i^2=1$, 
and the three angles $\phi_i$.

The complete supergravity solution 
(which is valid for generic complex values of $\beta$) can be found in the original paper \cite{LM}. 
In addition to the 
five-form field $F_5$ already present in the $AdS_5 \times S^5$ geometry,  the solution also includes 
the NS-NS two-form potential $B_2$ and  the RR potential $C_2.$

It is important to note that the dilaton $\phi$ is no longer 
constant, 
but depends on the coordinates $\mu_i$ of the deformed five-sphere. 
The constant parameter is
$\phi_0$ which has the meaning of the 
dilaton of the parent undeformed solution, and it maps to the coupling constant
of the dual gauge theory. However, it is $\phi$ and not $\phi_0$ which plays the r{\o}le of the
dilaton in the deformed supergravity solution.
The dilaton $\phi$ and the axion $C$ are assembled in the standard way into a complex $\tau$
\be
\tau \, =\, i e^{-\phi}\, + C 
\label{taucoup}
\ee
Equation \eqref{dildef} relates
this $\tau$ 
to the complexified coupling constant of the dual gauge theory, 
\be
\tau_0 \, =\, i e^{-\phi_0}\, + C \, =\, {4\pi i \over g^2} \, +\, {\theta \over 2\pi}
\label{tau0coup}
\ee
In summary,
the dictionary between the parameters of the deformed Yang-Mills theory and type IIB 
superstring theory on $AdS_5 \times S^5_{\gamma}$ is as follows:
\bea
e^{-\phi} \, G^{1/2}\, =\,  e^{-\phi_0} &=& {4\pi  \over g^2}
\, , \qquad C\,=\, {\theta \over 2\pi} \\
R^2 &=& \sqrt{g^2\,N}
\eea
and
$R$ is the radius of
the $AdS_5$ space in units of $\sqrt{\alpha'}.$
The supergravity background is a valid approximation to string theory 
in the small curvature regime \cite{LM}:
\be
R\, \gg \, 1 \ , \qquad R\betaR \, \ll\, 1
\ee
In terms of the gauge theory variables, the appropriate limit to consider is
\be
g^2 N \,\gg \,1 \ , \qquad N\, \gg\, 1 \ , \qquad \betaR\, \ll \, 1
\ee
In the above, the $N\gg 1$ condition arises from the fact that the $SL(2,Z)$ duality
can be used to map large string couplings to values which are not large.

As is well known, 
the supergravity action of type IIB theory is invariant under the 
non-compact symmetry group $SU(1,1) \sim SL(2,R)$. 
The action of this symmetry leaves the metric invariant,
but acts upon the dilaton-axion field $\tau$ of  Eq.~\eqref{taucoup}
\bea
\label{sl2zcoup}
\tau\longrightarrow \tau'=\frac{a\tau+b}{c\tau+d},\qquad ad-bc=1,\ a,b,c,d\in R.
\eea
The string theory is invariant only under the $SL(2,Z)$ subgroup of 
the $SL(2,R)$. This implies that the string theory effective
action $S_{\rm IIB}$ must be invariant under the $SL(2,Z)$
S-duality transformation.

The string effective action $S_{\rm IIB}$ is related via the
AdS/CFT holographic formula \cite{GKP,Witten150} to
correlation functions in the gauge theory,
\begin{equation}
 \exp-S_{\rm IIB}\left[\Phi_{\cal O};J\right]\, =\, 
 \left\langle \exp\,\int\, d^{4}x \, J(x){\cal O}(x)\,
\right \rangle
\label{corr}
\end{equation}  
Here $\Phi_{\cal O}$ are Kaluza-Klein modes of the supergravity fields 
which are dual to composite gauge theory operators $\cal O$.
The boundary conditions of the supergravity fields 
are set 
by the gauge theory sources on the boundary
of $AdS_5$ via $\Phi_{\cal O}(x) \, \propto\, J(x).$

Constructing all the Kaluza-Klein modes on the deformed five-sphere is a non-trivial task,
hence, in this paper
we are restricting ourselves to the 
lowest Kaluza-Klein modes which do not depend on the coordinates of the deformed sphere.

D-instanton contributions in supergravity arise
as $(\alpha^{\prime})^3$ corrections \cite{BG} 
to the classical IIB theory. The D-instanton contribution to
an $n$-point correlator $G_n$ comes from a tree level Feynman diagram
with one vertex located at a point $(x_0,\rho,\hat\Omega)$ in the bulk
of $AdS_{5}\times \tilde{S}^{5}$.  The diagram also has $n$ external legs
connecting the vertex to operator insertions on the boundary. There is a bulk-to-boundary
propagator associated with each external leg \cite{Witten150,FMMR}.
For example, an $SO(6)$ singlet scalar 
free field of mass $m$ on $AdS_{5}$ has the bulk-to-boundary propagator  
\begin{equation}
K_{\Delta}(x_0,\rho;x,0)=\frac{\rho^{\Delta}}{(\rho^{2}+(x-x_0)^{2})^{\Delta}}\, ,
\label{prop}
\end{equation}  
where $(mL)^2=\Delta(\Delta-4)$.
At leading order beyond the
Einstein-Hilbert term in the derivative expansion, the IIB effective
action is expected to contain \cite{GG1}, \cite{BG}
an ${\cal R}^4$ term\footnote{Here, ${\cal R}^4$ denotes 
a particular contraction \cite{GG1} of four ten-dimensional 
Riemann tensors.}
\be
(\alpha')^{-1} \int d^{10}x \,\sqrt{-g_{10}} \,e^{-\phi/2}
\, f_4(\tau,\bar\tau) \,
 {\cal R}^4 
\label{twoterms}
\ee
as well as its superpartners, including
a totally antisymmetric 16-dilatino
effective vertex of the form \cite{GGK,KP2}
\begin{equation}
(\alpha')^{-1}\int d^{10}x\,\sqrt{-g_{10}} \,e^{-\phi/2}\,
f_{16}(\tau,\bar\tau)\,\Lambda^{16}\ + \ \hbox{H.c.}
\label{effvert}\end{equation}
The dilatino $\Lambda$ is a complex chiral $SO(9,1)$ spinor which transforms
under the local  $U(1)$ symmetry with the charge $q_{\Lambda}=3/2$. 
Under the $SL(2,Z)$ transformations \eqref{sl2zcoup} all fields $\Phi$ are multiplied
by a (discrete) phase,
\be
\Phi\, \longrightarrow \, \left(\frac{c\tau+d}{c\bar\tau+d}\right)^{-\,q_{\Phi}/2} \Phi \ ,
\label{phases}
\ee
and the charge $q_{\Phi}$ for the dilatino is $3/2$ and for the ${\cal R}$ field it is zero.

Equations \eqref{twoterms}-\eqref{effvert}
are written in the string frame with the coefficients
$f_n(\tau,\bar\tau)$ being the modular forms of weights $((n-4),-(n-4))$
under the $SL(2,Z)$ transformations \eqref{sl2zcoup},
\be
f_n(\tau,\bar\tau) \, :=\,
f^{\,(n-4), -(n-4)} (\tau,\bar\tau) \, \longrightarrow \, \left(\frac{c\tau+d}{c\bar\tau+d}\right)^{n-4}
f^{\,(n-4), -(n-4)} (\tau,\bar\tau)\, .
\ee
The modular properties of $f_n$ precisely cancel the phases of fields in \eqref{phases}
acquired under the $SL(2,Z).$
Thus the full string effective action is invariant under the $SL(2,Z)$
and this modular symmetry 
ensures the S-duality of the type IIB superstring.

The modular forms $f_n$ have been constructed by Green and Gutperle 
in \cite{GG1}.
In the weak coupling expansion the expressions for $f_n$
contain an infinite sum of exponential terms
\begin{equation}
e^{-\phi/2}\, f_n \, \ni \,
\sum_{k=1}^{\infty}\, {\rm const}\cdot\left({k\over G^{1/2}\,g^2}\right)^{n-7/2}\, 
e^{2\pi ik\tau}\, \sum_{d|k}{1\over d^2}\ ,\label{foksh}
\end{equation}

In the original undeformed $\cN=4$ scenario, 
$\tau=\tau_0$ and the modular forms $f_n$ in the string effective action
can be thought of as functions of the gauge coupling constants $\tau_0$ and
$\bar\tau_0$. In this case, each of the terms in the expression above
must correspond 
to a contribution of an instanton of charge $k$.
On the other hand, the $k$-instanton contributions can be independently
calculated directly in gauge theory. These calculations have been performed
in \cite{BGKR,DKMV} at the 1-instanton level 
and in \cite{DHKMV,MO3} for the general $k$-instanton case.
Remarkably, the SYM results of \cite{BGKR,DKMV,DHKMV,MO3} 
tuned out to be
in precise agreement with the supergravity predictions
for the effective action Eqs.~\eqref{twoterms}-\eqref{foksh} and in 
Eq.~\eqref{fincorrs} below.

The goal of the present paper is to attempt to reproduce these results
in the $\beta$-deformed case. Hence we want to interpret 
the sum on the right hand side of \eqref{foksh} again as coming from 
multi-instantons in gauge theory. We see that, at least potentially,
there is a puzzle in this interpretation as the Yang-Mills $k$-instantons
are expected to contribute factors proportional to $e^{2\pi ik\tau_0}$ rather than
to $e^{2\pi ik\tau}.$ 
In the rest of the paper when we perform an explicit $k$-instanton calculation
in the $\beta$-deformed SYM we will find the resolution of this puzzle.

The main object of interest for us are the $n$-point correlation functions
of certain composite operators in the $\beta$-deformed SYM.
We can consider the same classes of the operators as in
\cite{BGKR,MO3,GK} which correspond to the lowest KK-modes in supergravity.
Specifically we will analyze the gauge-invariant chiral correlators
$G_n$, $n=16,$ 8 or 4, defined by:
\begin{subequations}\begin{align}
G_{16}\ &=\
\langle\,\cO(x_1) \cdots \cO(x_{16})\rangle\ ,\quad
\cO:=\, \Lambda_\alpha^A\ =\ g^{-2}\sigma^{mn}{}_\alpha^{\
\beta}\,{\rm tr}_N\,
F_{mn}\,\lambda_\beta^A\ ,
\label{e4}
\\
G_{8}\ &=\ \langle\,\cO(x_1)\cdots \cO(x_{8})\rangle\ ,\quad
\cO:=\, {\cal B}^{[AB]}_{mn} \ =\ 
g^{-2}\,{\rm tr}_N\, \big(\lambda^A \sigma_{mn}\lambda^B +
2i F_{mn}\Phi^{AB}
 \big)\ ,
\label{e66}
\\
G_{8}\ &=\ \langle\,\cO(x_1)\cdots \cO(x_{8})\rangle\ ,\quad
\cO:=\, {\cal E}^{(AB)} \ =\ 
g^{-2}\,{\rm tr}_N\, \big(\lambda^A \lambda^B +
t^{(AB)+}_{[abc]} \phi^a \phi^b \phi^c
 \big)\ ,
\label{e6}
\\
G_{4}\ &=\ \langle\,\cO(x_1)\cdots \cO(x_{4})\rangle\ ,\quad
\cO:=\,{\cal Q}^{ab} \ 
=\ g^{-2}\,{\rm tr}_N\,
\big(\phi^a \phi^b - \tfrac16\delta^{ab}\phi^c \phi^c \big)
\ ,
\label{e5}
\end{align}\end{subequations}
where $t$ in Eq.~\eqref{e6} is a numerical tensor.
In the notation of Ref.~\cite{GK} these correlators were called the minimal ones.
The non-minimal correlators $G_n$ with higher $n$
were considered in \cite{GK} in the context of the original $\cN=4$ AdS/CFT correspondence.
In the present paper we will concentrate on the minimal case above, and paying particular attention to 
the correlators in \eqref{e4} and \eqref{e66}.

The AdS/CFT holographic relation then predicts that these correlators 
must lead on the supergravity side to the following expressions:
\be
G_n \, \sim\, (\alpha')^{-1}\,\int d^4 x\, {d\rho\over \rho^5}\,
\int \,G\, d^5 \hat\Omega\,\, t_n\, e^{-\phi/2}\, f_n({\tau,\bar\tau}) \,
\prod_{i=1}^n K(x_0,\rho;x_i,0)
\label{fincorrs}
\ee
Here $G d^5 \hat\Omega$ represents the volume form on the $\tilde{S}^5$
and each $K$ denotes the bulk-to-boundary propagator which corresponds to each particular
operator in \eqref{e4}-\eqref{e5}.
Various index contractions between the $n$ states (propagators) are schematically represented
by a tensor $t_n$.
We want to verify the above relations using multi-instanton calculations
in the $\beta$-deformed SYM theory. As in Ref.~\cite{MO3} it will actually be sufficient
for this purpose to concentrate on the multi-instanton partition function.
The correlators can be obtained from the latter by inserting in it
the operators calculated on the instanton solution.

\section{Marginal $\beta$-deformations of $\cN=4$ SYM}

The $\beta$-deformed Yang-Mills is an $\cN=1$ supersymmetric conformal gauge theory
with a global $U(1)\times U(1)$ symmetry \eqref{u1s}.
Lunin and Maldacena have pointed out \cite{LM} that
the $\betaR$-deformation in \eqref{superpot2}
can be understood as arising from introducing the star products between the fields in the
$\N=4$ Lagrangian,
\be
\label{star}
f * g \equiv
e^{ i  \pi \betaR (Q_1^f Q_2^g - Q_2^f Q_1^g) } f g
\ee
Here $(Q_1^{\rm field},Q_2^{\rm field})$ are the $U(1)_1 \times U(1)_2$
charges of the fields ($f$ or $g$).
The values of the charges for component fields are read from \eqref{u1s}:
\bea
\Phi_1\ ,\ \lambda_1\,  : \,\,\,\, &&(Q_1\, , Q_2)=(0\, , -1)
\label{ch1}\\
\Phi_2\ , \ \lambda_2\,  : \,\,\,\, &&(Q_1\, , Q_2)=(1\, , 1)
\label{ch2}\\
\Phi_3\ , \ \lambda_3\,  : \,\,\,\, &&(Q_1\, , Q_2)=(-1\, , 0)
\label{ch3}\\
A_m \ , \ \lambda_4 \,  : \,\,\,\, &&(Q_1\, , Q_2)=(0\, , 0)
 \label{ch4}
\eea
and for the conjugate fields ($\bar\Phi_i$ and $\bar\lambda_i$) the charges are opposite.

The component Lagrangian of the $\betaR$-deformed theory
is easily read from the $\cN=4$ Lagrangian 
\bea
\nonumber
&&{\cal{L}} = \, \Tr \Bigg( {1 \over 4}
F^{\mu \nu}F_{\mu \nu} +
(D^\mu \bar \Phi^i ) (D_\mu \Phi_i  )
- {g^2\over 2} [\Phi_i,\Phi_j]_{*}[\bar \Phi^i,\bar \Phi^j]_{*}
+{g^2\over 4}[\Phi_i,\bar \Phi^i][\Phi_j,\bar \Phi^j] \\
&&+ \l_{A} \s^{\mu} D_{\mu} \bar\l^A
- i g ([\l_4,\l_i]\bar \Phi^i+[\bar\l^4,\bar \l^i]\Phi_i)
+{i g\over 2}(\epsilon^{ijk}[\l_i,\l_j]_{*}\Phi_k
+\epsilon_{ijk}[\bar\l^i,\bar\l^j]_{*}\bar \Phi^k)\Bigg)
\nonumber \\
\label{Ldef}
\eea
In the above equation the star products \eqref{star} are used
for fields charged under the $U(1)_1 \times U(1)_2.$
We have also used the fact that the star product is trivial between two
fields which have opposite $U(1)_1 \times U(1)_2$ charges.
We have also introduced the $\betaR$-deformed commutator of fields which is simply
\be
[ f_i,g_j]_{*} :=\, f_i * g_j - g_j * f_i = \,
e^{ i \pi \beta_{ij} }\, f_i g_j -\,
e^{ -i \pi \beta_{ij} }\, g_j f_i \ , \label{combeta}
\ee
and $\beta_{ij}$ is defined as
\be
\beta_{ij}=-\beta_{ji} \, , \quad
\beta_{12}=\, -\beta_{13}=\, \beta_{23} :=\, \betaR \ .
\label{betaijs}
\ee
More generally, and for future reference we also define
a $4 \times 4$ deformation matrix $\beta_{AB}$ with $A,B=1,\ldots,4$
\be
\beta_{AB}=-\beta_{BA} \, , \quad
\beta_{4i}=0 \, , \quad
\beta_{12}=\, -\beta_{13}=\, \beta_{23} :=\, \betaR \ .
\label{betaABs}
\ee
The component Lagrangian in the form \eqref{Ldef}
is well-suited for tracing the $\betaR$-dependence in perturbative
calculations and it was utilized in \cite{VVK}.

For carrying out multi-instanton calculations in the formalism of \cite{MO3,Review}
it is more convenient to switch to a different basis for scalar fields.
We assemble the three complex scalars $\Phi_i$ into an adjoint-valued
antisymmetric tensor field $\Phi^{AB}(x)$,
subject to
a specific reality condition:
\begin{equation}
\hf\epsilon_{ABCD}\, \Phi^{CD}\ =\ {\bar{\Phi}}^{AB}\, ,
\label{realcon}
\end{equation}
which implies that it transforms in the vector ${\bf 6}$
representation of $SO(6)_R$ symmetry of the $\cN=4$ SYM.
In terms of the six real scalars $\phi^a$
it can be written as \cite{MO3}
\begin{equation}
\Phi^{AB}=\frac1{\sqrt8}\bar\Sigma_a^{AB}\phi^a\ , \qquad
\bar{\Phi}^{AB}=\ - \, \frac1{\sqrt8}\Sigma_a^{AB}\phi^a\ , \qquad
a=1,\ldots,6 \label{Sigmaform}
\end{equation}
where the coefficients $\Sigma^{AB}_a$ and $\bar\Sigma^{AB}_a$ are expressed
in terms of the 't Hooft $\eta$-symbols:
\bea
\Sigma^a_{AB}&=&\big(\eta^1_{AB},i\bar\eta^1_{AB},\eta^2_{AB},i\bar\eta^2_{AB},
\eta^3_{AB},i\bar\eta^3_{AB}\big), \\
\bar\Sigma^{AB}_a&=&\big(-\eta^1_{AB},i\bar\eta^1_{AB},-\eta^2_{AB},i\bar\eta^2_{AB},
-\eta^3_{AB},i\bar\eta^3_{AB}\big)
\eea
Here $\eta$ and $\bar\eta$
are the selfdual and anti-selfdual 't Hooft symbols
\cite{TH}:
\begin{align}
&\bar\eta_{AB}^c=\eta_{AB}^c=\epsilon_{cAB}\qquad A,B\in\{1,2,3\}, \notag\\
&\bar\eta_{4A}^c=\eta_{A4}^c=\delta_{cA},\label{E15.1}\\
&\eta_{AB}^c=-\eta_{BA}^c,\qquad\bar\eta_{AB}^c=-\bar\eta_{BA}^c.\notag
\end{align}
The relation between the two bases of scalar fields is then given by
\bea
\Phi_1 = \, \frac{1}{\sqrt{2}} (\phi^1 +i\phi^2) &=& 2 \,\bar{\Phi}^{23} = \,2\,\Phi^{14}
\nonumber\\
\Phi_2 = \, \frac{1}{\sqrt{2}} (\phi^3 +i\phi^4) &=& 2 \,\bar{\Phi}^{31} = \,2\,\Phi^{24}
\label{dictphi}\\
\Phi_3 = \, \frac{1}{\sqrt{2}} (\phi^5 +i\phi^6) &=& 2 \,\bar{\Phi}^{12} = \,2\,\Phi^{34}
\nonumber
\eea
and their $U(1)_1 \times U(1)_2$ charges can be read off \eqref{ch1}-\eqref{ch3}

\section{Instantons in the $\beta$-deformed $\cN=4$ SYM}  \label{sec:insts}

In pure gauge theory, instantons obey the self-duality equation
\be
F_{mn} = {}^* F_{mn}
\label{sdeq}
\ee
The ADHM $k$-instanton \cite{ADHM,Corrigan,Christ} is the gauge configuration, $A_{m},$
which is the general solution of \eqref{sdeq}
with the topological charge $k$. When gauge fields are coupled to
fermions and scalars, as in the $\cN=4$ SYM, one needs to consider
the coupled classical Euler-Lagrange equations instead of \eqref{sdeq}.
Instanton configurations then also include fermion and scalar-field components.
Our goal, however, is not
just to find classical solutions, but rather to calculate their quantum
contributions to correlators $G_n$, which includes the effects of
a perturbative expansion in the instanton background.
The way to take the leading  perturbations into account
automatically is to modify the background configuration itself as
explained in \cite{MO3}; however, the instanton supermultiplet is then no longer an
exact solution to the coupled equations of motion.
In particular, $k$-instanton fermion components in the $\cN=4$ $SU(N)$ SYM
are defined \cite{MO3} to contain
all of the $8kN$ fermion zero modes of the Dirac operator,
$\Dbarslash^{\dalpha \alpha}$
and
not just the 16 exact unlifted zero modes. Similarly, in the $\beta$-deformed theory,
the same total of $8kN$ fermion zero modes will be included into the $k$-instanton supermultiplet,
even though only 4 of them are exact in the theory with $\cN=1$ supersymmetry.

As a result, our strategy
(see Section 2 of Ref.~\cite{MO3} and in Section 4 of Ref.~\cite{Review} for more detail)
is to solve Euler-Lagrange equations iteratively, order by
order in the Yang-Mills coupling.
In this paper we restrict our attention to the leading
semiclassical order, meaning the first non-vanishing order in $g$ at
each topological level. The relevant equations which define these
leading order instanton component
fields are the self-duality equation \eqref{sdeq}
together with the fermion zero-mode
equation
\be
\Dbarslash^{\dalpha \alpha} \lambda_{\alpha}^A =\ 0
\label{fzerms}
\ee
and the equation for the scalar field, which for the $\beta$-deformed theory takes the form
\be
\D^2 \Phi^{AB}\ =\ \sqrtwo
i\,(\,\lambda^A * \lambda^B\,-\,\lambda^B* \lambda^A\,)
\label{Higgseq}
\ee
Here $\Dbarslash^{\dalpha \alpha}= D^m \sigmabar_m^{\dalpha \alpha}$ and
$\D^2= D^m D_m$ where $D_m$ is the covariant derivative in the
instanton  background $A_m$. 

For convenience, in deriving classical equations above,
we have rescaled all fields in the Lagrangian with an overall factor of  $1/g$, so that
the only $g$ dependence in the action is through the overall
coefficient $1/g^2$. Hence, $g$ does not appear
on the right hand side of \eqref{Higgseq}.
The explicit $g$ dependence in the Euler-Lagrange
equations can be trivially restored by undoing this rescaling.

Equation \eqref{sdeq} specifies the gauge field instanton component $A_m$ of topological charge $k$.
The second equation \eqref{fzerms} defines gaugino components $\lambda^A$ of the instanton.
As already mentioned, all $8kN$ adjoint fermion zero mode solutions of \eqref{fzerms}
are included in the $k$-instanton supermultiplet. Only $4$ of these modes are protected by the
$\cN=1$ superconformal invariance of the theory and are exact zero modes. Remaining $8kN-4$
fermion zero modes will be lifted by interactions, this means that the instanton action
will depend on collective coordinates of these fermion modes.

The last equation \eqref{Higgseq} determines
the scalar field instanton components $\Phi^{AB}$ in terms of the gauge-field and gaugino components.
$\beta$-deformation affects only this equation and it appears via the star product
in the commutator on the right hand side. Apart from this obvious modification
in \eqref{Higgseq},
all three equations \eqref{sdeq}-\eqref{Higgseq} are the same as in the undeformed
$\cN=4$ SYM of Ref.~\cite{MO3}.

In order to evaluate correlators $G_n$ in the SYM picture, one inserts the $n$
appropriate gauge-invariant operators under the integration
$\int\dmuphys\,\exp(-\Skinst)$
where $\Skinst$ is the $k$-instanton action and $\dmuphys$ is the
collective coordinate measure.

The ADHM gauge-field and the gaugino components of the instanton
are parameterized by a set of collective coordinates.
The scalar field is entirely determined in terms of $A_m$ and $\lambda^A$
in \eqref{Higgseq} and
no new collective coordinates of the instanton are associated with $\Phi^{AB}$.
In general, there are $4kN$ {\it independent} bosonic
and $8kN$ {\it independent} fermionic collective coordinates of a $k$-instanton configuration
in our model.
The simplest way to define the collective coordinate integration measure $\dmuphys$
is to consider an even larger set of instanton collective coordinates
which are not all independent, but satisfy certain
algebraic equations -- the ADHM constraints \eqref{adhmcons}.

These bosonic and fermionic collective coordinates
live, respectively, in an
$(N+2k)\times2k$ complex matrix $a$, and in an $(N+2k)\times k$
Grassmann-valued complex matrix $\M^A,$ where the $SU(4)_R$ index
$A=1,2,3,4$ labels the \susy. In components:\footnote{We use notations
of Refs.~\cite{MO3,Review} throughout. The indices
$u,v=1,\ldots,N$ are $SU(N)$ indices;
$\alpha,\dot\alpha,\hbox{etc.}=1,2$ are Weyl indices (traced over with
`$\trtwo$'); $i,j=1,\ldots,k$ ($k$ being the topological number) are
instanton indices (traced over with `tr$_k$'); and $m,n=1,2,3,4$ are
Euclidean Lorentz indices. Pauli matrices are $(\tau^c)_\alpha^{\,\beta}$ where $c=1,2,3$.}
\begin{equation}
a\ =\
\begin{pmatrix}
w_{ui\dalpha}\\ \big(a'_{\beta\dalpha}\big)_{li}
\end{pmatrix}
\ ,\qquad
\M^A \ =\
\begin{pmatrix}
\mu^A_{ui}\\ \big(\M^{\prime A}_\beta\big)_{li}
\end{pmatrix}
\label{adef}\end{equation}
where both $a'_m$ (defined by
$a'_{\beta\dalpha}=a'_m\sigma^m_{\beta\dalpha}$)
and $\M^{\prime A}_\beta$ are Hermitian $k\times k$ matrices. These
matrices are subject to the ADHM constraints:
\be
\trtwo(\tau^c\abar a)_{ij}=\, 0 \ , \qquad
\big(\Mbar^Aa+\abar\M^A\big)_{\beta\, ij}=\, 0
\label{adhmcons}
\ee
as
well as to a $U(k)$ symmetry
\begin{equation}
w_{iu\aD}\rightarrow w_{ju\aD}U_{ji}\ ,\quad
a'_{mij}\rightarrow U^{-1}_{ik}\,a'_{mkl}\,U^{}_{lj}\ ,\quad U\in U(k)\ ,
\label{E28}\end{equation}

In the dilute instanton gas limit, the individual collective coordinates of the $k$
far-separated instantons are
\begin{subequations}
\begin{align}
x_n^i\ &=\ -(a'_n)_{ii}^{}\, ,
\label{xconmulti}\\
\rho_i^2\ &=\ \hf\wbar^\dalpha_{iu} \, w^{}_{ui \dalpha} \, ,
\label{wconmulti}\\
\left(t^c_i\right)_{uv}\  &=\ \rho_i^{-2}\,w^{}_{ui\aD}
\left(\tau^c\right)^\aD_{\ \bD}\bar
w_{iv}^\bD\, ,
\label{embedmulti}
\end{align}
\end{subequations}
where
$x_n^i$ are positions, $\rho_i$ are the sizes and
$t^c_i$ are the SU(2) generators of the individual instantons $i=1,\ldots,k.$

The gauge field, gaugino and scalar field components of the $k$-instanton configuration
which solve equations
\eqref{sdeq}-\eqref{Higgseq} can be found in Ref.~\cite{MO3,Review} for the
original $\cN=4$ SYM.
The multi-instanton components in the $\beta$-deformed theory are obtained from the
$\cN=4$ expressions in \cite{MO3,Review} simply
by applying the star products for all quantities charged under the $U(1) \times U(1)$ symmetry.

This multi-instanton configuration gives rise to the $k$-instanton action
\def\Skquad{S^k_{\rm quad}}
\begin{equation}
\Skinst\ =\ {8\pi^2k\over g^2}+\, ik\theta+\Skquad\ .
\label{E47}
\end{equation} 
Here $\Skquad$ is a term quadrilinear in
fermionic collective coordinates, with one fermion collective
coordinate chosen from each of the four gaugino sectors
$A=1,2,3,4\,$:
\begin{equation}
\Skquad\ =\ {\pi^2\over g^2}\,\epsilon_{ABCD} \, {\rm
tr}_k\left(\Lambda_{A*B}\, \BL^{-1}\,\Lambda_{C*D}\right).
\label{E48}
\end{equation}
The $k\times k$ anti-Hermitian fermion bilinear $\Lambda_{A*B}$ is
given by\footnote{The $*$-subscript in $\Lambda_{A*B}$ is used to
indicate that the right hand side of \eqref{Lambdadef} contains
the star-product of the Grassmann collective coordinates $\Mbar^A$
and $\Mbar^B$.}
\def\sqrtwo{\sqrt{2}\,}
\begin{equation}
\Lambda_{A*B}\ :=\ {1\over2\sqrtwo} \big(\Mbar^A * \M^B-\Mbar^B *
\M^A\big) \label{Lambdadef}
\end{equation}
and $\BL$ is a linear
self-adjoint operator that maps the $k^2$-dimensional space of
such matrices onto itself:
\def\hf{{\textstyle{1\over2}}}
\begin{equation}
\BL\cdot\Omega\ =\ \hf\{\Omega,\bar{w}^\dalpha 
w_\dalpha\}
+ \ [a_n',[\bar a_n',\Omega]]
\label{E35}
\end{equation}
The expression for the $k$-instanton action in the $\cN=4$ theory
was first derived in \cite{DKMn4} and subsequently used in
\cite{MO3,Review}. The only modification of this expression in the
$\beta$-deformed theory is the appearance of the star product
between the fermionic collective coordinates in \eqref{Lambdadef}.
In general, the Grassmann collective coordinates $\M^A$ and
$\Mbar^A$ are so far the only parameters appearing in the instanton
configuration which are charged under the $U(1) \times U(1)$.
Hence the $\beta$-dependence is recovered from the $\cN=4$ results
by introducing the star products in expressions involving $\M^A$
and $\Mbar^A$ parameters.\footnote{ We note that $\Lambda_{A*B}$
and $\epsilon_{ABCD}\,\Lambda_{C*D}$ are oppositely charged under
the $U(1) \times U(1)$ and hence the star product is not needed on
the right hand side of \eqref{E48}. } 
In the original $\cN=4$ theory, $\Skquad$ lifts all
the adjoint fermion modes except the 16 exact \susic\ and
superconformal modes. The $\beta$-deformed theory 
lifts $8kN-4$ fermion modes. The unlifted modes are the two
supersymmetric $\lambda_{\rm ss}^{\alpha \, A=4}$ and two
superconformal modes $\lambda_{{\rm sc}\, \dalpha}^{ \, \, A=4}$
of the unbroken $\cN=1$ supersymmetry.

Following Ref.~\cite{MO3}, we now want to simplify $\Skquad$ 
by integrating in some new
bosonic parameters. The idea is to replace the fermion
quadrilinear $\Skquad$ with a fermion bilinear coupled to a
 set of auxiliary Gaussian variables. These take
the form of an anti-symmetric tensor $\chi_{AB}=-\chi_{BA}$ whose elements are
$k\times k$ matrices in instanton indices. We have
\begin{equation}
\exp\left(-\Skquad\right)\  =\
\pi^{-3k^2}\,
\big({\det}_{k^2}\BL\big)^{3}
  \int
d^{6k^2}\chi\exp\big[\,-{\rm tr}_k\left(\epsilon_{ABCD}\, \chi_{AB}\,\BL\,
\chi_{CD}\right)+4\pi ig^{-1}{\rm
tr}_k\left(\chi_{AB}\,\Lambda_{A*B}\right)\,\big]
\label{E53}
\end{equation}
The variable $\chi_{AB}$ transforms in the vector
representation of the $\SO(6)\cong\SU(4)$ R-symmetry and is subject to
the reality condition 
$\chi_{AB}^\dagger = \, \hf
\epsilon^{}_{ABCD}\chi^{}_{CD}$

Next we turn to the $k$-instanton $\N=4$ collective coordinate
measure. This measure involves integrations over all bosonic and
fermionic collective coordinates of the $k$-instanton,
subject to the bosonic and fermionic ADHM constraints \eqref{adhmcons}.
These constraints are implemented via insertions of the appropriate delta functions~\cite{DHKM}.
The $k$-instanton integration
measure for $\cN=4$ SYM reads \cite{MO3}:
\def\wbar{\bar w}
\def\mubar{\bar\mu}
\def\abar{\bar a}
\begin{equation}
\begin{split}
\int \dmuphys\ =&\ {2^{-k^2/2}(C_1)^k\over{\rm Vol}\,U(k)}
\int d^{4k^2} a' \
d^{2kN} \wbar \ d^{2kN} w  
\prod_{A=1}^{4}
\ d^{2k^2} \M^{\prime A}  
\ d^{kN} \mubar^A \ d^{kN} \mu^A\\
&\times 
({\rm det}_{k^2} \BL)^{-3}
\prod_{r=1}^{k^2}\Big[
\prod_{c=1,2,3}
\delta\big(\tfrac12{\rm
tr}_k\,T^r(\trtwo\, \tau^c \abar a)\big)
\prod_{A=1}^{4}
\prod_{\aD=1,2}\delta\left({\rm tr}_k\,T^r(\Mbar^A a_\aD + \abar_\aD \M^A
)\right)\Big]\ .
\label{measuredef}
\end{split}
\end{equation}
The constant $C_1$ is fixed at the 1-instanton level \cite{MO3}
\begin{equation}
C_1=2^{-2N+1/2}\pi^{-6N}g^{4N}
\end{equation}
by
comparing Eq.~\eqref{measuredef} with the standard 1-instanton 't Hooft-Bernard
measure \cite{TH,Bernard}.
The
integrals over the $k\times k$ matrices $a'_n$, $\CM^{\prime A}$ and
${\cal A}^{AB}$ are defined in \eqref{measuredef}
as the integral over the components with
respect to a Hermitian basis of $k\times  k$ matrices $T^r$ normalized
so that ${\rm tr}_k\,T^rT^s=\delta^{rs}$.

The complete instanton partition function, $\int\dmuphys\,\exp(-\Skinst),$
in the $\beta$-deformed gauge theory is given by\footnote{Note that 
\eqref{partf} does not contain
factors of $\det\BL$; they cancelled out between Eqs.~\eqref{E53}
and \eqref{measuredef}.
}
\begin{equation}
\begin{split}
&{(C_1)^k 2^{-k^2/2}\pi^{-3k^2}\over{\rm Vol}\,U(k)}
\int d^{4k^2} a' \
d^{2kN} \wbar \ d^{2kN} w  \ d^{6k^2}\chi \, 
\prod_{A=1}^{4}
\ d^{2k^2} \M^{\prime A}  
\ d^{kN} \mubar^A \ d^{kN} \mu^A\\
& \prod_{r=1}^{k^2}\Big[
\prod_{c=1,2,3}
\delta\big(\tfrac12{\rm
tr}_k\,T^r(\trtwo\, \tau^c \abar a)\big)
\prod_{A=1}^{4}
\prod_{\aD=1,2}\delta\left({\rm tr}_k\,T^r(\Mbar^A a_\aD + \abar_\aD \M^A
)\right)
 \\
&\exp\big[\,-{8 \pi^2 \over g^2}-{\rm tr}_k\left(\epsilon_{ABCD}\, \chi_{AB}\,\BL\,
\chi_{CD}\right)+ {4\pi i \over g} {\rm tr}_k\left(\chi_{AB}\Lambda_{A*B}\right)\,\big]
\end{split}
\label{partf}
\end{equation}
We note that
this expression differs from the original $\cN=4$ result of \cite{MO3} only 
through the star product in the last term in the exponent. In the following Section we will
see that this fact has profound consequences.  

In the Appendix we also give an alternative string theory derivation  of  \eqref{partf}.
We show there that our gauge theory expression \eqref{partf}
is identical to the partition function of $k$ D-instantons in string theory with
the $\beta$-deformation.


\section{The large-$N$ saddle-point integration: 1-instanton case} \label{sec:one-inst}

We will first carry out integrations over collective coordinates in the simplest
case of a single instanton $k=1$. The generalization to multi-instantons will be discussed
in the following Section.

One way to carry out the single-instanton calculation, is to first solve the ADHM constraints
\eqref{adhmcons}, and then to integrate out the 
corresponding delta-functions in \eqref{partf}.
The collective coordinates
\eqref{adef} which satisfy the $k=1$ ADHM constraints
can be written in the simple canonical 
form \cite{KMS,DKMV}:
\be
\label{matdef}
a\ =\ \begin{pmatrix}
0\quad 0 \\
\vdots\quad \vdots\\
0\quad 0\\
\rho
\quad 
0\\
0\quad \rho\\
-x_0^m\sigma_m
\end{pmatrix}
\ ,\quad
\M^A\ =\ \begin{pmatrix}
\nu^A_1\\
\vdots\\
\nu^A_{N-2}\\
4i\rho\etabar^{A1}\\
 4i\rho\etabar^{A2}\\
 4\xi_1^A\cr4\xi_2^A \end{pmatrix}
\ee
Here we follow the usual notation where $\rho\in\bigR$ and $x_0^m\in
\bigR^4$ denote the size and position of the instanton, 
and $\xi_\alpha^A$
and $\etabar^{A\dot\alpha}$ are the supersymmetric and superconformal
fermion zero modes, respectively.
Equation \eqref{matdef} assumes the
canonical `North pole' embedding of the $SU(2)$ instanton within $SU(N)$;
more generally there is a manifold of equivalent instantons obtained
by acting on \eqref{matdef} by transformations $\Omega$ in the coset space
$\Omega\ \in\ U(N)/ (U(N-2)\times U(1)).$
The complex Grassmann coordinates $\nu_i^A$ in
Eq.~\eqref{matdef} (which do not
carry a Weyl spinor index) may be thought of as the superpartners of the coset
embedding parameters $\Omega$.
 Together, $\xi_\alpha^A,$ $\etabar_\dalpha^A,$
$\nu_i^A$ and $\nubar_i^A$ constitute $8N$ fermionic 
collective coordinates of a single instanton in the $\cN=4$ SYM.

After integrating out the delta functions imposing the ADHM constraints,
and integrating over the instanton iso-orientations $\Omega$,
the instanton measure \eqref{partf} for $k=1$ reduces to:
\bea
\int d\mu_{\rm phys}^1\,e^{-S_{\rm inst}^1} \, =\,
{2^{-31}\pi^{-4N-5} g^{4N}\over(N-1)!(N-2)!}
\,\int d^4 x_0\, d\rho \, d^6 \chi \, 
\prod_{A=1}^{4} d^2\xi^A\, d^2\bar\eta^A\,
d^{(N-2)}\nu^A\, d^{(N-2)}\bar\nu^A
\nonumber \\
\rho^{4N-7}\, \exp\big[-{8\pi^2\over g^2} +i\theta- 2\rho^2 \chi^a \chi^a
+{4\pi i \over g}\,\chi_{AB}\Lambda_{A*B}\big]
\label{1instpf}
\eea
The integral is expressed in terms of the collective coordinates
\eqref{matdef} and we have substituted the 1-instanton expression  
$\BL = 2\rho^2$.
The fermion bilinear in the 1-instanton case reads
\be
\Lambda_{A*B} = \, {1 \over 2\sqrt{2}}\, \sum_{i=1}^{N-2}\left(e^{i\pi \beta_{AB}}\,\bar\nu^A_i \nu^B_i 
-e^{-i\pi \beta_{AB}}\,\bar\nu^B_i \nu^A_i\right) 
+ i 8\sqrt{2}\, \sin(\pi \beta_{AB}) \,\left(\rho^2 \bar\eta^A \cdot \bar\eta^B +\xi^A \cdot\xi^B\right)
\label{Lam1inst}
\ee
In the expression above we used the fact the products of Grassmann parameters with 
the Weyl index $\xi^A \cdot \xi^B := \xi^{A\alpha} \xi^B_{\alpha}$ and 
$\bar\eta^A \cdot \bar\eta^B :=\bar\eta^A_{\dot\alpha} \bar\eta^{B\dot\alpha}$
are symmetric in $A$ and $B$. The $4 \times 4$ antisymmetric matrix
$\beta_{AB}$ was defined in \eqref{betaABs}.

It is worth noting that the second term on the right hand side of \eqref{Lam1inst}
is non-vanishing only for non-zero values of the deformation parameter $\beta_{AB}$.
This implies that there are precisely four exact fermion zero modes
in the $\beta$-deformed theory
which do not enter \eqref{Lam1inst}:
two supersymmetric ones, $\xi^4_\alpha,$ and two superconformal $\bar\eta^4_{\dot\alpha}$ modes.
In the undeformed $\cN=4$ theory, all 16 supersymmetric and superconformal modes 
were absent from $\Lambda_{AB}$ and hence from the instanton action.

However, even though only 4 out of 16 supersymmetric and superconformal modes
are exact, they will altogether be irrelevant for our purposes. The main point
here is the fact that we can choose such correlators that
all 16 of these modes will be saturated by the instanton expressions for the Yang-Mills operators
inserted into the partition function \eqref{1instpf}.
At the same time we require that all of the remaining $\nu$ and $\bar\nu$ modes in the instanton
partition function should not be lifted by the insertions of the 
operators. For this to be correct, we first of all need to restrict ourselves to the 
insertions of gauge invariant composite operators
which correspond to zero KK modes on the deformed sphere $\tilde{S}^5$ in supergravity.\footnote{Insertions 
of the operators corresponding to non-zero KK modes would lift some of the
$\nu$ and $\bar\nu$ modes, as in the $\beta=0$ case studied in \cite{MO3,GK}.}
Secondly, we have to restrict to the minimal 
correlators \eqref{e4}-\eqref{e5} of these operators. In the case of $\langle\Lambda^{16}\rangle$ correlator \eqref{e4}
and the $\langle{\cal B}^{8}\rangle$ correlator \eqref{e66} all of the 16 supersymmetric/superconformal modes and
none of the  $\nu$ and $\bar\nu$ modes are lifted by the operator insertions.\footnote{For the non-minimal correlators
involving higher values of $n$ this is not the case anymore. At the same time, even the minimal correlators
$\langle{\cal E}^{8}\rangle$ and $\langle{\cal Q}^{4}\rangle$
in \eqref{e6}-\eqref{e5} can receive corrections from saturating some of the $\nu$ and $\bar\nu$ modes
by the operator insertions. This would then require one to keep 
(part or all) of the 12 lifted supersymmetric/superconformal
modes in the exponent.
These corrections can in principle be straightforwardly calculated in the small $\beta$-limit, which is the regime
relevant for comparison with the supergravity. We thank Stefano Kovacs for pointing this out to us.
For more detail on the fermion-zero-mode structure of the operator insertions we refer the reader to 
Ref.~\cite{GK}.}
In summary, for our purposes of calculating the minimal correlators in \eqref{e4}, \eqref{e66} (and also
in \eqref{e6}, \eqref{e5} in the small-$\beta$ regime)
one can always neglect the second term on the right hand side of \eqref{Lam1inst},
which is what we will do from now on.

 We can now start integrating out fermionic collective coordinates
 $\nu_i^A$ and $\bar\nu_i^A$ from the instanton partition function \eqref{1instpf}.
 For each value of $i=1,\ldots,N-2$ this integration gives a factor of 
 \be
 \left({4 \pi \over g}{1 \over \sqrt{2}}\right)^4 \, 
 {\rm det}_4 \left(e^{i\pi \beta_{AB}}\,\chi_{AB}\right)
 \label{detbchi}
\ee
The determinant above can be calculated directly. It will be useful
to express the result in terms of the three complex variables
$X_i$ which are defined in terms of $\chi_{AB}$ in the way analogous to
Eqs.~\eqref{dictphi}:
\bea
X_1 = \, \chi^1 +i\chi^2 &=& 2\sqrt{2} \,{\chi}^\dagger_{23} = \,2\sqrt{2}\,\chi_{14}
\nonumber\\
X_2 = \, \chi^3 +i\chi^4 &=& 2\sqrt{2} \,{\chi}^\dagger_{31} = \,2\sqrt{2}\,\chi_{24}
\label{dictchi}\\
X_3 = \,  \chi^5 +i\chi^6 &=& 2\sqrt{2} \,{\chi}^\dagger_{12} = \,2\sqrt{2}\,\chi_{34}
\nonumber
\eea
In terms of these degrees of freedom, the $\beta$-deformed determinant takes the form
\bea
&&{\rm det}_4 \left(e^{i\pi \beta_{AB}}\,\chi_{AB}\right) \, = 
\label{defdet} \\
&&{1 \over 64} \, \left(
(|X_1|^2 + |X_2|^2 + |X_3|^2)^2  \,- \, 4\, \sin^2 (\pi \beta_R)\,
(|X_1|^2|X_2|^2 + |X_2|^2|X_3|^2 + |X_1|^2|X_3|^2)\right) \nonumber
\eea
It follows that the determinant depends only on the three absolute
values of $|X|$ and is independent of the three angles. 
We can further change variables as follows:
\be
\label{mudefs}
|X_i| =\, r\, \mu_i \ , \qquad \sum_{i=1}^3 \mu_i^2 =\,  1
\ee
and write
\be
 \left({4 \pi \over g}{1 \over \sqrt{2}}\right)^4 \, 
 {\rm det}_4 \left(e^{i\pi \beta_{AB}}\,\chi_{AB}\right)=\, 
  \left({\pi \over g}\right)^4 \, r^4 \, \left(1\, -\, 
  4\, \sin^2 (\pi \beta_R)\, Q\right) 
  \label{e58}
\ee
where
\be 
\label{Qmudefs}
Q :=\, \mu_1^2\mu_2^2 + \mu_2^2\mu_3^2 + \mu_1^2\mu_3^2
\ee
We also split the integral over $d^6 \chi$ in the partition function
into an integral over $r$ and the integral over the 5-sphere $\hat\Omega$:
\be 
\label{chisplit}
\int d^6 \chi = \, \int r^5\, dr \, d^5 \hat\Omega \ , \qquad {\rm where} \qquad
\int d^5 \hat\Omega \propto \, \int d\mu_1^2 \,d\mu_2^2 \,d\mu_3^2 \, \,
\delta(\sum_{i=1}^3 \mu_i^2 - 1)
\ee
In summary after integrating out all of the $\nu$ and $\bar\nu$ fermionic
collective coordinates we find
\bea
\int d\mu_{\rm phys}^1\,e^{-S_{\rm inst}^1}\ \ = \
{g^8 \over2^{31}\pi^{13}(N-1)!(N-2)!}
\int d^4 x_0\,{d\rho \over \rho^5}\,d^5\hat\Omega\,
\prod_{A=1,2,3,4}
d^2\xi^A\, d^2\bar\eta^A 
\nonumber\\
e^{-{8\pi^2\over g^2}+i\theta}\,\left(1\, -\, 4\, \sin^2 (\pi \beta_R)\, Q\right)^{N-2}\,\,
\rho^{4N-2}\,I_N
\label{OneIM}
\eea
Here $I_N$ denotes the $r$ integration, which it is instructive to
separate out:
\begin{equation}
I_N=\int_0^\infty\, dr\,r^{4N-3}\,e^{-2\rho^2r^2}=
\tfrac12(2\rho^2)^{-2N+1}\int_0^\infty\,dx\,x^{2N-2}\,e^{-x}
=
\tfrac12(2\rho^2)^{1-2N}(2N-2)!
\label{INdef1}
\end{equation}

{}From Eqs.~\eqref{OneIM}-\eqref{INdef1} one sees that the $x_0$ and $\rho$
integrations assemble into the scale-invariant $AdS_5$ volume form
$d^4 x_0\,d\rho\,\rho^{-5}$ and also the integration over the 
5-sphere arises from $d^5\hat\Omega$.
However, it also follows that the final result given by 
Eqs.~\eqref{OneIM}-\eqref{INdef1} is so far unsatisfactory from the perspective of the
supergravity interpretation. 
First of all, the $\chi$ variables gave rise to the integration over the 
{\it undeformed} sphere $S^5$. Secondly, the $\betaR$-dependent factor, 
$\sin^2 (\pi \beta_R)\, Q,$  is neither spherically-invariant (due to its $Q$-dependence)
nor can it be easily associated with the deformed sphere $\tilde{S}^5$. Finally,
the exponent of the instanton action in \eqref{INdef1} is of the form 
$e^{-8\pi^2 \over g^2+i\theta} = e^{2 \pi i \tau_0}$ which is not of the form
$e^{2 \pi i \tau}$ expected in supergravity.

Quite remarkably, all of these perceived problems of Eqs.~\eqref{OneIM}-\eqref{INdef1}
can be resolved by taking the large-$N$ limit and carefully specifying 
the appropriate order of limits procedure. We will now describe this procedure in detail.

On the gauge theory side we have no choice but to work in the weak-coupling limit.
To justify working with the leading-order instanton and neglecting 
and infinite set of higher order terms
in perturbation theory in the instanton background, we must take the limit
$g^2 N \to 0$ and only after that impose the large $N$ limit.
In addition, so far we have been treating the $\beta$-deformation parameter 
as an independent fixed constant. However, 
as we have seen in Section {\bf 2}, the validity of the Lunin--Maldacena
supergravity solution is restricted to the regime of small $\beta$.
We proceed with the Yang-Mills instanton calculation by taking the limits in the following order:
\begin{enumerate}
\item $\qquad g^2 \ll 1 \qquad$ with $N$ and $\betaR$ being fixed
\item $\qquad N \gg 1 \ ,\quad \betaR^2 \ll 1$
\end{enumerate}
The second limit shall be taken in such a way that $\betaR^2 N $ is held fixed.
In fact, our Yang-Mills results are more general than that and hold for any values of 
the parameter $\betaR^2 N.$ It can be shown that the derivation below holds for
$\betaR^2 N \gg 1$ and $\betaR^2 N \ll 1.$

We now consider the factor
\be
{\cal F} := \, e^{-{8\pi^2\over g^2}+i\theta}\,\left(1\, -\, 4\, \sin^2 (\pi \beta_R)\, Q\right)^{N-2}
\label{Fdef1}
\ee
which appears on the right hand side of \eqref{OneIM}
and simplify it according to the limits above. First, we replace
the $\sin^2 (\pi \beta_R)$ 
 by $(\pi \beta_R)^2$ which is justified since $\betaR^2 \ll 1$.
 Next, we note that in our limit
 \be
 \left(1\, -\, 4\, (\pi \beta_R)^2\, Q\right)^{N-2} \, \sim \,
 \exp \left[ N\, \log(1- 4\pi^2 \betaR^2 \, Q) \right] 
 \, \sim \, \exp \left[- N \betaR^2 \,4\pi^2  \, Q \right]
 \ee
In the last expression above we have neglected
the higher-order terms ${\cal O}(N \betaR^4 Q^2) \sim 1/N  \ll 1 $ 
which arose from the expansion of the
logarithm. 

This allows us to write down the ${\cal F}$-factor defined in 
\eqref{Fdef1} as
\be
{\cal F} \,=\, \exp \left[-{8\pi^2\over g^2}\, +\, i\theta \,-\, N\betaR^2\, 4\pi^2\, Q\right]
\,=\, \exp \left[-{8\pi^2\over g^2}\left(1 \,+\, \hf\,N\betaR^2 g^2\, Q\right)\,+\, i\theta\right]
\label{Fdef2}
\ee
We stress that the expression above is exact in the ordered 
weak-coupling--large-$N$--small-$\beta$ limit which we use in our semi-classical
Yang-Mills calculation.

We now compare this ${\cal F}$-factor arising from the Yang-Mills instanton in the 
weak coupling limit to $e^{2\pi i \tau}$ in the Lunin-Maldacena
supergravity solution.
We recall that $\tau$ is the parameter which combines the natural dilaton and the axion of the deformed
supergravity solution
\be
\tau \, =\, i e^{-\phi} \,+\, C\,:=\, i\tau_2 \,+\, \tau_1
\ee
This $\tau$ is related to  $\tau_0$ and hence to the SYM couplings 
via a non-trivial function $G$ as follows\footnote{The axion $C$ or the Yang-Mills
$\theta$ parameter are not changed by this transformation and the real parts
of the two $\tau$'s are the same $\tau_{1}=\tau_{01}.$
When working with instantons we will not pay attention to the $\theta$ parameter,
if required it can always be trivially restored in the instanton action. }
\be
\label{lmsugexp}
e^{-2\pi  \tau_2} \, = \, e^{-2\pi \tau_{02} \, G^{-1/2}} \ , \qquad 
\tau_0 \, =\, {2\pi i \over g^2} + {\theta \over 2\pi}\, :=\,
 i\tau_{02} \,+\, \tau_{01}
\ee 
Here $G$ is the function of the coordinates $\mu_i$ on the deformed sphere, 
and it also
depends on the deformation parameter $\hat\gamma$
\be
G^{-1} \, =\, 1 \, + \, {\hat\gamma}^2
(\mu_1^2\mu_2^2 + \mu_2^2\mu_3^2 + \mu_1^2\mu_3^2) \ ,
\qquad 
{\hat\gamma}^2 \,=\, \betaR^2 N g^2
\label{LMGdef}
\ee

We now want to take the same
weak-coupling--large-$N$--small-$\beta$ limit
of the supergravity expressions \eqref{lmsugexp}-\eqref{LMGdef}.
We find
\be
e^{-2\pi \tau_2}
\, = \,
\exp \left[-{8\pi^2\over g^2}\left(1 \,+\,N\betaR^2 g^2\, Q\right)^{1 \over 2}\right]
\, = \,
\exp \left[-{8\pi^2\over g^2}\, -\, 4\pi^2 \,N\betaR^2\, Q \,
+\, \sim\, (N\betaR^2)^2 g^2 Q^2 \, + \, \ldots \right]
\ee
The linear in $Q$ term is unaffected in the limit, while
the quadratic term is negligible, $(N\betaR^2)^2 g^2 Q^2 \, \sim\, g^2 \, \ll 1$.
This gives
\be
e^{2\pi i \tau} \, = \, {\cal  F} \ .
\label{1instequiv}
\ee

{}From these considerations we conclude that the ${\cal F}$-factor which arises from the Yang-Mills
instanton calculation in the semi-classical limit of the deformed theory
is equivalent to the corresponding supergravity factor $e^{2\pi i \tau}$
of Eq.~\eqref{lmsugexp} in the Lunin-Maldacena background.
For this equivalence it is necessary to identify the $\mu_i$ coordinates of the 
instanton $\chi$-collective-coordinates defined in \eqref{mudefs}, \eqref{Qmudefs}
with the $\mu_i$ coordinates of the deformed $\tilde{S}^5$ sphere of the Lunin-Maldacena
background.
This implies that the integration measure over the `angles of' $\chi_a$, or more precisely over the
the 5-dimensional manifold $\hat\Omega$ in \eqref{chisplit}
should correspond to the volume element on the deformed $\tilde{S}^5$ sphere. 
This volume element $\omega_{\tilde{S}^5}$ over the deformed sphere
$\tilde{S}^5$ can be found from 
the Lunin-Maldacena metric. In the string frame we get
\be 
\label{voltils5}
\int \omega_{\tilde{S}^5} \, = \, \int \omega_{S^5} G
\ee
where $\omega_{S^5}$ is the volume element of the original 5-sphere,
and $G$ is given in \eqref{LMGdef}.

We have seen above that when $G^{-1/2}$ appears in the exponent  weighted with the
instanton action, it gives rise to two terms in the semiclassical limit:
the order-1 term and the order-$Q$ term.
However, when $G$ appears in the pre-exponent, as on the right hand side of
\eqref{voltils5}, it is indistinguishable from unity.
Indeed,
\be 
G \,=\, (1\,+\, g^2 N \betaR\,Q)^{-1} \, =\,
1 \, -\, g^2 N \betaR^2\,Q \, +\, \ldots \, \to \, 1
\ee
since 
$  g^2 N \betaR^2\,Q \,\sim\, g^2 \, \ll\, 1$ and should be neglected.
This amounts to identifying
\be 
\label{voltils51}
\int \omega_{\tilde{S}^5} \, = \, \int d^5 \hat\Omega\,  G \, \to \, \int d^5 \hat\Omega
\ee

For consistency
we can also re-calculate $I_N$ in \eqref{OneIM} in the large-$N$ limit.
First
one rescales $r\rightarrow\sqrt Nr$, or equivalently,
$ \chi^a\rightarrow\sqrt N\chi^a,$
so that
$N$ factors out of the exponent.  The integral then becomes
\begin{equation}
I_N=N^{2N-1}
\int_0^\infty\, dr\,r^{-3}\,e^{2N(\log r^2-\rho^2r^2)}\ ,
\end{equation}
which is in a form amenable to a standard saddle-point evaluation. 
The saddle-point is at $r=\rho^{-1}$ and, to leading order, a Gaussian
integral around the solution gives
\begin{equation}
\lim_{N\rightarrow\infty}\,I_N\ =\
\rho^{2-4N}N^{2N-1}e^{-2N}\sqrt{\pi\over N}\ , 
\end{equation}
which is valid up to $1/N$ corrections.

Our final result for the single-instanton partition function in the semi-classical large-$N$ limit
takes the following
simple form:
\bea
\int d\mu_{\rm phys}^1\,e^{-S_{\rm inst}^1}\ &\to&
{\sqrt Ng^8\over
2^{33}\pi^{27/2}}\,
\int\,
{d^4 x_0\,d\rho\over\rho^5}\, d^5\hat\Omega\prod_{A=1,2,3,4}d^2\xi^A 
d^2\bar\eta^A\ e^{-2\pi  \tau_{02} G^{-1/2}+2\pi\tau_{1}}
\nonumber\\
&=&
{\sqrt Ng^8\over
2^{33}\pi^{27/2}}\,
\int\,
d^{10}X \, \sqrt{-g_{10}} \, \prod_{A=1,2,3,4}d^2\xi^A 
d^2\bar\eta^A\ e^{2\pi i \tau}
\label{1instendexp}
\eea
where the integration over $d^{10}X \, \sqrt{-g_{10}}$ is the integration of the 10-dimensional space 
which corresponds to the Lunin-Maldacena background. 
This 10-dimensional bosonic integration can be factored into the ${AdS_5}$-part
parameterized by the instanton position and the scale-size,
$d^4 x_0\,{d\rho\over\rho^5},$ times the 5-dimensional integration over the deformed sphere
$\tilde{S}^5$ parameterized by  $d^5\hat\Omega \, G\,\to \, d^5\hat\Omega.$
The main feature of our 1-instanton result \eqref{1instendexp} is the appearance of the
complete dilaton-axion factor in the exponent, $e^{2\pi i \tau}.$


\section{Multi-instanton large-$N$ integration} \label{sec:large-N}

In this Section we return to the general case of $k$ instantons. Following
the saddle-point approach of \cite{MO3} and also building  upon the 1-instanton
calculation of the previous Section, we will evaluate the
multi-instanton partition function \eqref{partf}.

The first step is to reduce the $k$-instanton measure to the 
$SU(N)$-gauge-invariant expression \cite{MO3}.
The expression
\eqref{partf} can
be simplified by transforming to a smaller set of
gauge-invariant collective coordinates (i.e., variables without an
uncontracted `$u$' index). In the bosonic sector this means changing
variables from $\{w,\wbar\}$ to the $W$ variables
introduced as follows:
\begin{equation}
\big(W_{\ \bD}^\aD\big)_{ij}=\bar w_{iu}^\aD \,w_{ju\bD}\ ,\quad
W^0={\rm tr}_2\left(W\right),\quad W^c={\rm
tr}_2\left(\tau^cW\right), \ \ c=1,2,3\ .\label{E26}
\end{equation}
This enables us to reduce the number of bosonic integrations
using the Jacobian identity is proved in \cite{MO3}:
\begin{equation}
d^{2Nk}\wbar\,d^{2Nk}w\ =\ c_{k,N}\,\big(\det_{2k}W\big)^{N-2k}\,
d^{k^2}W^0\prod_{c=1,2,3}
d^{k^2}W^c\ ,
\label{interesting}\end{equation}
where $c_{k,N}$ is a constant. 
An important feature of this change of variables is that it allows us
to eliminate the bosonic ADHM constraints \cite{DHKM,MO3}.
The
ADHM constraints in Eq.~\eqref{partf}, which are quadratic in the
$\{w,\wbar\}$ coordinates, become linear in terms of $W$
\def\sigmabar{\bar\sigma}
\def\etabar{\bar\eta}
\def\zetabar{\bar\zeta}
\def\mubar{\bar\mu}
\def\nubar{\bar\nu}
\begin{equation}
0\ =\ W^c \ + \ [\,a'_n\,,\,a'_m\,]\,\trtwo
(\tau^c\sigmabar^{nm})\ =\
 W^c \ - \ i \ [\,a'_n\,,\,a'_m\,]\,\etabar^c_{nm}\ .
\label{E27}\end{equation}
We therefore use Eq.~\eqref{E27} to eliminate $W^c$ from the measure
together with the delta-functions of the bosonic
ADHM constraints.
Furthermore, we note that as $N\rightarrow\infty,$ the Jacobian factor of
$(\det W)^N=\exp(N\,\tr\log W)$ in \eqref{interesting} will be amenable 
to a saddle-point treatment.

In the fermion sector, following \cite{DHKM,MO3},
we change variables from
$\{\mu,\mubar\}$ to $\{\zeta,\zetabar,\nu,\nubar\}$ defined by
\begin{equation}
\mu_{iu}^A=w_{uj\aD}\,\zeta^{\aD A}_{ji}+\nu_{iu}^A\ ,\qquad
\bar\mu_{iu}^A=\bar\zeta_{\aD ij}^A\,\bar w_{ju}^\aD+\bar\nu_{iu}^A\ ,
\label{E44}\end{equation}
where the $\nu$ modes form a basis for the $\perp$-space of $w\,$:
\begin{equation}
0\ = \ \bar w_{iu}^\aD\nu_{ju}^A\ =\  \bar\nu_{iu}^Aw^{}_{ju\aD}\ ,
\label{E45}\end{equation}
In these variables the fermionic ADHM constraints in
Eq.~\eqref{partf} have the gauge-invariant form
\begin{equation}
0\ =\ \zetabar^A\,W+W\,\zeta^A+[\M^{\prime A},a']
\label{E46}\end{equation}
which can
 be used to eliminate
 $\bar\zeta^A$ in favor of  $\zeta^A$ and ${\cal M}^{\prime A}$;
doing so  gives a factor which precisely cancels
the Jacobian for the change of variables \eqref{E44}.

Since the
 $\nu$ and $\nubar$ modes are absent from  the constraint
\eqref{E46}, they can now be straightforwardly integrated out.
First, we decompose
 $\Lambda_{A*B}=\hat\Lambda_{A*B}+\tilde\Lambda_{A*B}$,
where
\begin{equation}
(\hat\Lambda_{A*B})_{ij}={1\over2\sqrt2}\left(\bar\nu_{iu}^A
*\nu_{ju}^B-
\bar\nu_{iu}^B*\nu_{ju}^A\right)\ ,
\label{E51}\end{equation}
and
\begin{equation}
\tilde\Lambda_{A*B}={1\over2\sqrt2}\left(\
\bar\zeta^A *W\zeta^{B}-\bar\zeta^B* W\zeta^{A}
+
[{\cal M}^{\prime A}\,,\,{\cal M}^{\prime B}]_*\ \right)\ .
\label{E52}\end{equation}
Second, we calculate
\begin{equation}
\int d^{4k(N-2k)}\nu\,d^{4k(N-2k)}\bar\nu\,\exp\left({4\pi i\over
g}\,{\rm tr}_k(\chi_{AB}\hat\Lambda_{A*B})\right)=
\left({8\pi^2\over g^2}\right)^{2k(N-2k)} 
\left({\rm det}^{}_{4k}\,e^{i\pi\beta_{AB}} \chi_{AB}\right)^{N-2k}
\label{E55}\end{equation}
To simplify notation we introduce a notation $q_{AB}= e^{i\pi\beta_{AB}}$, so the 
determinant in \eqref{E55} can be written as\footnote{Note, 
that $q\chi$ is a shorthand for 
$q_{AB} \chi_{AB}$ which is a product of matrix elements and not the product 
of two matrices.}
$({\rm det}^{}_{4k}\,q\chi)^N$. 
It too will contribute to the saddle-point
equations in the large-$N$ limit,
similarly to the $({\rm det} W)^N$ factor in Eq.~\eqref{interesting}.
The third and final contribution to
these equations will be the Gaussian term $\chi \BL \chi$ in Eq.~\eqref{partf},
once one rescales
 $\chi_{AB}\rightarrow\sqrt N\chi_{AB}$ so that $N$ factors out in
front. 
Combining the above manipulations, we write down the final expression for the
$SU(N)$-gauge-invariant measure (cf. \cite{MO3}):
\begin{equation}\begin{split}
\int\dmuphys\,e^{-\Skinst}\ =\ &
 {g^{8k^2}N^{k^2}e^{-8\pi^2k/g^2+ik\theta}
\over2^{27k^2/2-k/2}\,\pi^{13k^2}\,{\rm Vol}(\U(k))}\int d^{k^2}W^0\,d^{4k^2}a'\,
d^{6k^2}\chi\,\prod_{A=1,2,3,4}d^{2k^2}\M^{\prime A}\,
d^{2k^2}\zeta^A
\\
&\times\ \left({\det^{}_{2k} W}\,{\det^{}_{4k}q\chi}\right)^{-2k}\,
\exp\big[-\ {N S^{k\,(\beta)}_{\rm eff}} \,+\,
4\pi i g^{-1}\sqrt{N}\,{\rm tr}_k(\chi_{AB}\tilde \Lambda_{A*B})
\big]
\label{E57.2}
\end{split}\end{equation}
The constant in front of the integral is written in the large-$N$ limit.
In the exponent in the last line of \eqref{E57.2} we have 
grouped all of the order-$N$ terms into the quantity $N S^{k\,(\beta)}_{\rm eff}.$
The quantity $S^{k\,(\beta)}_{\rm eff}$
is the sum of the three
terms relevant for the large-$N$ saddle point approach 
mentioned above plus a constant piece
\begin{equation} 
S^{k\,(\beta)}_{\rm eff} \,:=\,
-\tr_{2k}\log W
\,-\,\tr_{4k}\log q\chi 
\,+\,\epsilon_{ABCD}\,{\rm tr}_k\left(\chi_{AB}\,
\BL\,\chi_{CD}\right)\,-\, 2k\left(1+3\log2\right)
\label{E57}\end{equation}
This expression
involves the $11k^2$ bosonic variables comprising the eleven
independent $k\times k$ Hermitian matrices $W^0$, $a'_n$ and
$\chi_a$. As mentioned earlier, the remaining components $W^c$, $c=1,2,3,$ are eliminated
in favor of the $a'_n$ via the ADHM constraint.  The
action is also invariant under the $\U(k)$ symmetry which
acts by adjoint action on all the variables.

We can apply the large-$N$ saddle-point formalism to the integral
in \eqref{E57.2}, but before doing so we want to slightly simplify $S^{k\,(0)}_{\rm eff}$
with respect to its $\betaR$-dependence. Specifically, we consider the
$\tr_{4k}\log q\chi$ term in $S^{k\,(\beta)}_{\rm eff}$ and split the $U(k)$ variables 
$\chi_{ij}$ into the sum of the $U(1)$ variables $\chi^{\star} \delta_{ij}$
and the $SU(k)$ degrees of freedom $\hat\chi_{ij}$
\be
\chi_{Ai\,Bj} \, =\, \chi^{\star}_{AB} \delta_{ij}\, +\, \hat\chi_{Ai\,Bj}\ , 
\qquad \tr_k\, \hat\chi_{AB}\, =\,0
\ee
We then expand the $\tr_{4k}\log q\chi$ as follows
\be
S^{k\,(\beta)}_{\rm eff}
 \, \in \, N\,\tr_{4k}\log q\chi\,=\,
 Nk\,\tr_{4}\log q\chi^\star\, + \,
 N\,\tr_{4k}\log\,\left(1+(q\chi^\star)^{-1}(q\hat\chi)\right) 
\ee 
and further re-write it as
\be
Nk\,\tr_{4}(\log q\chi^\star\,-\, \log \chi^\star)\,  + \,
 N\,\tr_{4k}\log\,\left(\chi^\star+\chi^\star(q\chi^\star)^{-1}(q\hat\chi)\right)
 \label{dqchisi}
 \ee
We can now take the small-$\betaR$ limit (accompanied by the large-$N$ limit).
The first term on the right hand side in \eqref{dqchisi} is equal to $k$ times
the single instanton result derived in the previous Section,
which is $k$ times $N \betaR^2\,  4\pi^4\, Q.$ 
Hence, for this term, the relevant contribution
comes at the order-$N\betaR^2$
in the $N \to \infty,$ $\betaR\to 0$ limit. 

However, a careful fluctuations analysis along the lines of \cite{MO3},
shows that in the second term in \eqref{dqchisi} the dominant contribution
comes at the order $N(\betaR)^0\sim N$. This makes all higher-order terms in $\betaR$
suppressed in the limit.
Thus we set $q=1$ in the second term which then reads:
\be
N\,\tr_{4k}\log\,\left(\chi^\star+\hat\chi\right) \, =\, N\,\tr_{4k}\log \chi
\ee

In summary, we express
\be
NS^{k\,(\beta)}_{\rm eff} \,=\, -\,Nk\,\tr_{4}(\log q\chi^\star\,-\, \log \chi^\star)\,  + \,
S^{k\,(0)}_{\rm eff}
\label{Seffsplit}
\ee
where $S^{k\,(0)}_{\rm eff}$ is given by \eqref{E57} with the substitution $\beta=0$ 
or equivalently $q=1.$
\def\Skeff{S^{k\,(0)}_{\rm eff}}
The first term in \eqref{Seffsplit} is combined with the $k$-instanton action
$8\pi^2 k/g^2+ik\theta$ in exactly the same way as in Eqs.~\eqref{Fdef2}, \eqref{1instendexp}
in the previous Section. This amounts to promoting the 
Yang-Mills multi-instanton gauge action to the appropriate $\tau$-dependence
required in the supergravity effective action
\be
\exp \left[-{8\pi^2 k \over g^2} +ik\theta \,-\, N k \betaR^2\, 4\pi^2\, Q\right] \, =\,
e^{-2\pi k \tau_{02} G^{-1/2}+2\pi k \tau_{1}} \, =\, e^{2\pi i k \tau}
\ee

What remains is $S^{k\,(0)}_{\rm eff},$ which does not depend on the deformation
parameter, it is the same as in the $\cN=4$ SYM
 theory and is amendable to
the large-$N$ saddle-point treatment.
The saddle-point approach has been set up and the integrations 
around the saddle-point solution have been carried
out in \cite{MO3}. Here we will only give a brief summary of the result. 
It turns out that the dominant contribution to the integral
comes from the maximally degenerate saddle-point solution:
\begin{equation}
W^0=2\rho^2\,1_{\sst[k]\times[k]},\qquad
\chi_a=\rho^{-1}\hat\Omega_a\,1_{\sst[k]\times[k]},\qquad
a'_n=-x_n\,1_{\sst[k]\times[k]}\ ,
\label{specsol}
\end{equation}
which corresponds to $k$
coincident Yang-Mills instantons of the same scale-size $\rho$ which live in the
mutually commuting $SU(2)$ subgroups of the $SU(N).$
In the supergravity interpretation this saddle-point corresponds to a 
configuration
living at a common point $\{x_n,\hat\Omega_a,\rho\}$ in the deformed 
$AdS_5\times \tilde{S}^5.$ This is a point-like object -- the D-instanton of charge $k.$

Around the special solution, the bosonic fluctuations fall into three
sets. First, there are $10$ zero modes which correspond to the
position of the $k$-instanton
``bound state'' in  $AdS_5\times \tilde{S}^5$. These are exactly the same as in the 
1-instanton case.
Second, there
are $k^2$ fluctuations called $\varphi$
which have a nonzero quadratic coefficient in the small-fluctuations expansion.
The remaining $10k^2-10$
fluctuations first appear beyond quadratic order and they correspond to
the traceless i.e. $SU(k)$ parts 
$\hat\chi_{AB}$, and $\hat a'_m$ of the ten $k\times k$ matrices
$\chi_{AB}$ and
$a'_m$. 
Since fluctuations over $\varphi$ are 
Gaussian, they can be straightforwardly integrated out.

To complete the
expansion, we include the fermion
terms in the exponent of \eqref{E57.2}.
The second term in the exponent involves fermionic degrees
of freedom appearing in $\tilde \Lambda_{A*B}$. Here we
again are interested in the
leading order non-vanishing contributions in the $\betaR \to 0$ limit. 
This amounts to dropping the star product 
$\tilde \Lambda_{A*B}\,\to\, \tilde \Lambda_{AB}.$
The resulting fermion
terms in the exponent involve
the traceless parts $\hat\zeta^{\aD A}$ and $\hat{\cal
M}^{\prime A}_\alpha$ coupled to $\hat a'_m$ and $\hat\chi_{AB}.$ 

Remarkably, in the large-$N$ limit,
the leading-order terms of the effective action
around the saddle-point solution, with the quadratic fluctuations
$\varphi$ integrated out, precisely assemble themselves into the
dimensional reduction from ten to zero of ${\cal
N}=1$ supersymmetric Yang-Mills with gauge group $\SU(k)$ in {\it
flat\/} space. The $\SU(k)$ adjoint-valued ten-dimensional gauge field and
Majorana-Weyl fermion are defined in terms of the fluctuations:
\begin{equation}
A_\mu=N^{1/4}\left(\rho^{-1}\hat a'_m,\rho\hat\chi^a\right),\qquad
\Psi=\
\Big({\pi\over2g}\Big)^{1/2}\,N^{1/8}\left(\rho^{-1/2}\hat{\cal
M}^{\prime A}_\alpha,\rho^{1/2}\hat\zeta^{\aD A}\right).
\end{equation}
The action for the dimensionally reduced gauge theory is
\begin{equation}
S(A_\mu,\Psi)=-{1\over2}{\rm tr}_k\,\left[A_\mu,A_\nu\right]^2\
+\ {\rm
tr}_k\left(\bar\Psi\Gamma_\mu\left[A_\mu,\Psi\right]\right).
\end{equation}

We conclude that the effective gauge-invariant
measure for $k$ instantons in
the large-$N$ limit
 factorizes into a 1-instanton-like piece, for the position
of the bound state in $AdS_5\times S^5$ and the $16$
supersymmetric and superconformal modes, times the partition function
${\cal Z}_k$ of the
dimensionally-reduced ${\cal N}=1$ supersymmetric $\SU(k)$  gauge theory
in flat space:
\begin{equation}\begin{split}
\int\dmuphys\,e^{-\Skinst}\ 
 {=}\ &
{\sqrt Ng^8\over k^3 2^{17k^2/2 -k/2
+25}\,\pi^{9k^2/2+9}}\\
&\times\ \int
{d\rho \over \rho^5}\,d^4x\, d\Omega_5\prod_{A=1,2,3,4}d^2\xi^A
d^2\bar\eta^A \, e^{-8\pi^2k/g^2}\, \hat{\cal Z}_k\ ,
\label{hello}\end{split}\end{equation}
where $\hat{\cal Z}_k$ is the partition function of an ${\cal N}=1$
supersymmetric $\SU(k)$ gauge theory in ten dimensions dimensionally reduced
to zero dimensions:
\begin{equation}
\hat{\cal Z}_k\ =\ {1\over{\rm Vol}\,SU(k)}\int_{SU(k)}\, 
d^{10}A\, d^{16}\Psi\,e^{-S(A_\mu,\Psi)}\ .
\label{sukpart}
\end{equation}
Notice that the rest of the measure, up to numerical factors, is
independent of the instanton number $k$.
When integrating expressions which are independent of the $\SU(k)$
degrees-of-freedom, $\hat{\cal Z}_k$ is simply an overall constant
factor. A calculation of Ref.~\cite{MNS,KNS} revealed that $\hat{\cal
Z}_k$ is proportional to $\sum_{d|k}d^{-2}$, a sum over the positive integer
divisors $d$ of $k$. In our notation we have \cite{MNS,KNS,MO3}:
\begin{equation}
\hat{\cal Z}_k=2^{17k^2/2-k/2-8}\pi^{9k^2/2-9/2}k^{-1/2}
\sum_{d\vert k}{1\over d^2}\ .
\label{parte}\end{equation}

In summary, on gauge invariant and $\SU(k)$ singlet operators, our
effective large-$N$ collective coordinate measure has the following
simple form:
\begin{equation}
\int\dmuphys\,e^{-\Skinst}\ 
{=}\ {\sqrt{Ng^2} \over
2^{33}\pi^{27/2}}\,{k\over g^2}^{-7/2}\sum_{d\vert k}{1\over d^2}
\int\,
{d^4x\,d\rho\over\rho^5}\, d^5\hat\Omega \prod_{A=1,2,3,4}d^2\xi^A 
d^2\bar\eta^A\\ e^{2\pi ik \tau} \ .
\label{endexp}
\end{equation}
We can already identify  a number of key features of the $k$-instanton measure
which are important for the comparison with the supergravity results 
\eqref{twoterms}-\eqref{foksh}.
First, the factor of $(k/ g^2)^{-7/2}$ in the measure maps nicely to the
factor in $(k/ g^2)^{n-7/2}$  for $n=0$ on the right hand side of \eqref{foksh}.
Second, we recognize the inverse divisors squared contributions $\sum_{d\vert k}{1\over d^2}$
in \eqref{endexp} and \eqref{foksh}. The matching of $e^{2\pi ik \tau}$ factors 
has been mentioned earlier and it is one of our main results. The factor of $\sqrt{Ng^2}$ in \eqref{endexp}
gives rise to $(\alpha')^{-1}$ in \eqref{twoterms}, and the volume element of the
$AdS_5$ is represented via ${d^4x\,d\rho\,/\rho^5}$ in \eqref{endexp}.
Finally, the integration over $d^5\hat\Omega$ gives rise to the volume factor of the
5-sphere. However, as we have already explained earlier, in our semi-classical
limit we cannot distinguish between the deformed and the undeformed 
spheres {\it in the pre-exponent}.
The deformation is, however, manifest in the exponential factor
$e^{2\pi ik \tau}.$

\section{Correlation functions}

Finally, we can  use our measure to calculate the correlation functions
$G_n(x_1,\ldots,x_n)$ listed in \eqref{e4}-\eqref{e5}.
This entails inserting into Eq.~\eqref{endexp}
the appropriate product of gauge-invariant composite chiral operators
$\cO_1(x_1)\times\cdots\times\cO_n(x_n)$, which together contain
the requisite 16 exact fermion modes to saturate the 16 Grassmann integrations\footnote{We
also refer the reader to the earlier discussion following Eq.~\eqref{Lam1inst} and to footnote 9.} 
in \eqref{endexp}.
Since,
at leading order in $N$, the $k$  instantons sit
at the same point in $\AdS_5\times \tilde{S}^5$, it follows that
$\cO_j^{(k)}$ is simply
proportional to its single-instanton counterpart:
$\cO_j^{(k)}=k\cO_j^{(1)}$.
Therefore $G_n$ scales like ${(k/g^2)}^n$. This promotes the factor
in the partition function to the full value required to match with \eqref{foksh}
\be
\left({k\over g^2}\right)^{-7/2} \, \longrightarrow \, \left({k\over g^2}\right)^{n-7/2}
\ee
and, as before, factors of $G$ in the pre-exponent in \eqref{foksh}
cannot be tested in our limit.

Furthermore, it was shown in \cite{BGKR,MO3} that the instanton contributions
to the operators $\cO$ precisely match the functional form
of the bulk-to-boundary propagators in \eqref{fincorrs}.
We thus conclude that our Yang-Mills multi-instanton results for the correlators
$G_n$ which follow from Eq.~\eqref{endexp}
completely reconstruct the supergravity expressions
Eqs.~\eqref{twoterms}-\eqref{foksh},\eqref{fincorrs}.

\bigskip

As the final comment we recall that the matching between the supergravity and the SYM results 
holds in the opposite limits.
The SYM expression is derived in the weak coupling limit $g^2 N \to 0,$ $N\to \infty$
while the supergravity is a good approximation to string theory in the strong coupling
limit $g^2 N \to \infty,$ $N\to \infty.$ 
This use of different limits on the two sides of the AdS/CFT correspondence 
is, of course, the consequence of the strong-to-weak coupling nature of the AdS/CFT.
Nevertheless, even though the two sets of limits are mutually exclusive, we have shown that
the leading order results in the SYM and in supergravity agree with each other.
This agreement between the strong and the weak coupling limits holds in the instanton case 
(as it did hold in the original $\cN=4$ settings in \cite{BGKR,DKMV,DHKMV,MO3}), but
it is not expected to hold in perturbation theory. 
At present no non-renormalization theorem is known which would apply
to these instanton effects and explain the agreement. However the fact that there
is an agreement between the results on the two sides of the correspondence
must imply a non-trivial consistency of the AdS/CFT. We refer the reader to 
Refs.~\cite{MO3,GoGreen} for a more detailed discussion on this point.

\section{Complex $\beta$ deformations} \label{sec:gencompl}

We now consider the more general case 
of marginal deformations with complex values of the deformation parameters.
We will first explain how to extend the instanton calculation on the gauge theory
side from real to complex $\beta$-deformations. We will carry out this calculation
for arbitrary (not necessarily small) values of the deformation parameter $\beta \in C$.
We will also encounter
an interesting finite renormalization of the gauge coupling $\tau_0$
which has been predicted in \cite{DHK}. One of the main results 
is the instanton prediction for 
the dilaton-axion field $\tau$. We will show that in the limit of small $\beta$
it will match precisely with the $\tau$ field of the Lunin-Maldacena supergravity 
dual \cite{LM}. The small-$\beta$ limit is required \cite{LM} to ensure the 
validity of the supergravity approximation to full string theory. At the same time,
the SYM instanton calculation is valid for any finite values of complex $\beta$
and (as always) it is valid to the leading order at weak coupling $g^2 \ll 1$, 
and the large-$N$ limit.

The gauge theory description follows from the superpotential
\be
\label{superpotC}
i h \,\Tr( e^{ i \pi \beta } \Phi_1 \Phi_2 \Phi_3 - e^{-i \pi \beta } \Phi_1 \Phi_3 \Phi_2 )\ ,
\ee
where $h$ and $\beta$ are two complex parameters, and we will also use
$\beta$,
\be
\beta\, =\, \betaR \, +\, i\, \betaIm
\ee
Leigh and Strassler \cite{LS} have made an important observation 
that the superpotential \eqref{superpotC} gives an
exactly marginal deformation of the $\cN=4$ SYM theory if the three complex parameters,
$\tau_0$, $h$ and $\beta$ satisfy a single constraint, $\gamma(\tau_0,h,\beta)=0$.
We have already mentioned that for real values of $\beta$ this constraint can be solved exactly in the large-$N$ limit,
and amounts to $h=g$ with $\beta=\betaR$ and arbitrary.  
For complex deformations, 
the exact form of the function $\gamma$ in the conformal constraint is not 
known. In principle, one can solve the constraint in perturbation theory. To the leading order 
in $g^2 \ll 1$ and $N \ll 1$ it gives
\be
|h|^2 \, {\rm cosh}(2\pi \betaIm)\ =\, g^2
\ee
 see e.g. \cite{FG,MPSZ,VVK}.
It will turn out, however, that for our calculation
we will not need to know the explicit resolution of the 
constraint.

As before,
in writing down classical equations we always rescale all fields with an overall factor of  $1/g$,
so that the action goes as $1/g^2$.
The instanton configuration at the leading order in weak coupling is defined
by equations \eqref{sdeq}-\eqref{Higgseq}, with the scalar field equation \eqref{Higgseq}
being slightly modified as
\bea
&\D^2 \Phi^{AB}\ =\ {h\over g}\, \sqrtwo
i\, (\,e^{i\pi\beta^{AB}}\, \lambda^A  \lambda^B\,-\,e^{-i\pi\beta^{AB}}\,\lambda^B \lambda^A\,)
\qquad &{\rm for} \, A,B \neq 4 \ ,
\label{Higgseq2}\\
&\D^2 \Phi^{AB}\ =\  \sqrtwo
i\,(\,\lambda^A  \lambda^B\,-\,\lambda^B \lambda^A\,)
\qquad &{\rm for} \, A \, {\rm or}\, B = 4
\label{Higgseq4}
\eea
The factor of $h/g$ on the right hand side of \eqref{Higgseq2} accounts for the change of
the coupling constant from $g$ in \eqref{superpot2} to $h$ in \eqref{superpotC}.
Deformation parameters $h$ and $\beta$ are complex
and  $\beta^{AB}= {\betaR}^{AB} +i \betaIm^{AB}$.

We note that the resulting instanton configuration depends on $h$ holomorphically,
i.e. it does not depend on $h^*$. At leading order in $g$ the dependence on $h^*$ 
can come only through the equation conjugate to \eqref{Higgseq2}, which involves 
anti-fermion zero modes $\bar\lambda$ on the right hand side. These are vanishing 
in the instanton background. 
It is clear then that the anti-instanton configuration, will depend on $h^*$ and not on $h$.

Equations \eqref{Higgseq2}-\eqref{Higgseq4} imply that
the fermion quadrilinear term \eqref{E48} in the instanton action acquires an additional
factor of $h/g$
\begin{equation}
\Skquad\ =\ {\pi^2\over g^2}\,{h \over g}\, \epsilon_{ABCD} \, {\rm
tr}_k\left(\Lambda_{AB}\, \BL^{-1}\,\Lambda_{CD}\right)
\label{E48h}
\end{equation}
where $\Lambda_{AB}$ is given by (cf. \eqref{Lambdadef})
\def\sqrtwo{\sqrt{2}\,}
\begin{equation}
\Lambda_{AB}\ =\ {1\over2\sqrtwo} \big(\,e^{i\pi\beta^{AB}}\,\Mbar^A  \M^B\, -
\,e^{-i\pi\beta^{AB}}\,\Mbar^B 
\M^A\big) \label{Lambdadefh}
\end{equation}

Following the approach of Section {\bf 5} we now concentrate on the 1-instanton sector and
bilinearize the quadrilinear term \eqref{E48h} by introducing collective coordinates $\chi_{AB}$.
We then can integrate out fermionic collective coordinates
$\nu_i^A$ and $\bar\nu_i^A$.
For each value of $i=1,\ldots,N-2$ this integration gives a factor of the
determinant \eqref{detbchi} times an appropriate rescaling by $h/g$. The rule \eqref{E48h} 
is that there
is one power of $h/g$ for each factor of $1/g^2$
In total we have
\be
 \left({1 \over g}\right)^4 \, 
 {\rm det}_4 \left(e^{i\pi \beta^{AB}}\,\chi_{AB}\right) \, \longrightarrow\,
  \left({1 \over g}\right)^4 \, \left({h \over g}\right)^2\,
 {\rm det}_4 \left(e^{i\pi \beta^{AB}}\,\chi_{AB}\right) \ .
  \label{e58h}
\ee
This determinant can be evaluated as in Eq. \eqref{e58}.
The resulting characteristic instanton factor is
\be 
{\cal F} \, =\, e^{-{8\pi^2 \over g^2}+i\theta} \, {\left [
\left({h \over g}\right)^2\,\left(1\, -\, 4Q\, \sin^2 (\pi \beta)\right) \right ]}^{N-2}
\label{Fcomp1}
\ee
where $\beta$ and $h$ are complex parameters and the function $Q$ is the same as in \eqref{Qmudefs}.
In the large-$N$ limit we can write
\be
{\cal F} \, =\, \exp\left[ 2\pi i \tau_0\, +\, 2N\log \left({h \over g}\right)\,
+\, N \log \left(1\, -\, 4Q\, \sin^2 (\pi \beta)\right) \right ]
\label{Fcomp2}
\ee

Dorey, Hollowood and Kumar \cite{DHK} have argued that the certain combinations
of parameters in the $\beta$-deformed gauge theory must transform as modular forms
under the action of $SL(2,Z)_s$. More precisely, \cite{DHK}
\be
\tau_r\, \longrightarrow\, {a\tau_r +b \over c\tau_r+d} \ , \qquad
\beta\, \longrightarrow\, {\beta \over c\tau_r+d} \ , \qquad
(h/g)^2 \sin(\pi \beta)\, \longrightarrow \, {(h/g)^2 \sin(\pi \beta) \over c\tau_r+d} \ .
\label{sl2zpars}
\ee
Here the parameter $\tau_r$ is obtained from the complexified gauge coupling 
$\tau_0 = 4\pi i /g^2 +\theta/(2\pi)$ by a finite shift (or renormalization),
\be
\tau_r \, :=\, \tau_0 \, -\, {i N \over \pi} \, \log {h \over g} \ .
\label{tauRdef}
\ee
It is pleasing to note that the  first two terms on the right hand side of our
instanton prediction \eqref{Fcomp2} assemble precisely into $2 \pi i \tau_r$,
\be
{\cal F} \, =\, \exp\left[ 2\pi i \tau_r\, 
+\, N \log \left(1\, -\, 4Q\, \sin^2 (\pi \beta)\right) \right ]
\label{Fcomp3}
\ee
If we now take the small deformation limit, $\betaR \ll 1$, $\betaIm \ll 1$,
appropriate for comparison with the Lunin-Maldacena solution, we find\footnote{
We would like to stress that even in the small deformation limit,
the difference between the parameters $\tau_r$ and $\tau_0$ is significant.
This difference is given by the second term on the right hand side of Eq.~\eqref{Fcomp2}.
In the weak-coupling limit one can resolve the conformal constraint and approximate this term via
$2N\log \left({h / g}\right)\,\sim\, -\,N \log\left({\rm cosh}(2\pi \betaIm)\right) \, \sim\,
-\,2\pi^2 \, N \betaIm^2$.
This contribution is not small in the large-$N$ limit.}                                                            
\bea
{\cal F} =\, \exp \left[2\pi i \tau_r \,-\,4\pi^2\,N Q\,
(\betaR^2-\betaIm^2+2i\betaR \betaIm)\right]
\label{Fcom}
\eea

To compare the SYM instanton prediction of Eq.~\eqref{Fcomp3} or Eq.~\eqref{Fcom}
to the supergravity contribution ,
${\cal F} = e^{2\pi i \tau}$, we need the expressions for the dilaton $\phi$ and the axion $C$
fields
in the Lunin-Maldacena background \cite{LM} for complex $\beta$. 
The Lunin-Maldacena supergravity solution is given in terms of fields
which depend on the filed theory parameters $\beta$ and $\tau_r$. Here it is important to stress that
that these parameters must be those which transform as modular forms under the 
$SL(2,Z)_s$ in \eqref{sl2zpars}. Hence the supergravity dual depends on $\tau_r$ rather than on $\tau_0$.
It is convenient to represent the shifted coupling $\tau_r$ of Eq.~\eqref{tauRdef}
as 
\be
\tau_r \, =\, {4\pi i \over g^2_r} \, +\, {\theta_r \over 2\pi} \, =\, 
i e^{-\phi_0} \, + \, C^0 
\label{tauRrepr}
\ee
The last equality defines the parameters $e^{-\phi_0}$ and $C^0$ appearing in the supergravity solution.
They should be distinguished from the imaginary and real parts of the original complexified gauge coupling
\be 
\tau_0 \, =\, {4\pi i \over g^2} \, +\, {\theta \over 2\pi}.
\ee

The dilaton and axion field components of the Lunin-Maldacena supergravity dual \cite{LM}
are given by
\bea
&e^\phi \, = \, e^{\phi_0}\,G^{1/2}H \, ,\qquad 
&C\, =\, C ^0 \,- \, {\hat\gamma}{\hat\kappa}\,e^{-\phi_0} \,H^{-1}Q \, ,
\\
&G^{-1}\, = \, 1 \,+ \,({\hat\gamma}^2+{\hat\kappa}^2)\, Q  ~,\qquad
&H \,=\, 1 \,+\, {\hat\kappa}^2\, Q       \ ,
\eea
where the function $Q$ is the same as previously and, similarly to the real case, we have defined 
\be
\hat\gamma\, :=\, \betaR\, g_r \sqrt{N} \ , \qquad
\hat\kappa\, :=\, \betaIm\, g_r \sqrt{N} \ . \label{hatdefs}
\ee

We now compare the characteristic exponential factor 
\eqref{Fcom} arising from the Yang-Mill instanton, to 
the exponential $e^{2\pi i \tau}$  
expected in the modular form terms in the IIB effective action
in the Lunin-Maldacena background.
We have
\bea
e^{2 \pi i \tau}\, =\, e^{2 \pi i (i e^{-\phi}+C)}\, =\, 
\exp  \left[-2 \pi e^{-\phi_0}[1+\hf({\hat\gamma}^2-{\hat\kappa}^2)Q]\, 
+\, 2\pi i(C^0 - e^{-\phi_0}{\hat\gamma}{\hat\kappa} Q)\right]
\eea
In the second equality we have used the expressions for the dilaton and the 
axion fields of the $\beta$-deformed background in the weak-coupling large-$N$
 small-$\beta$ limit, i.e. in the expansion in terms of ${\hat\kappa}$
 and ${\hat\gamma}$ we ignore terms of order cubic or higher.
By employing the relations \eqref{hatdefs} for the hatted parameters, we arrive at
\bea
e^{2 \pi i \tau}\, =\, \exp  \left[-{8\pi^2\over g_r^2}+i\theta_r \,
-\, 4\pi^2\,N Q\,(\betaR^2-\betaIm^2)\,  -\,i\, 8 \pi^2\, NQ\, \betaR \betaIm  \right],
\eea
By   comparing the last equation to 
\eqref{Fcom} it is immediate to see that 
\be
{\cal F}=e^{2 \pi i \tau}\ .
\ee
This provides a detailed and a non-trivial test of both,
the Lunin-Maldacena supergravity solution for complex $\beta$-deformations, and the 
expected structure of the string theory effective action.

\bigskip
\bigskip

\centerline{\bf Acknowledgements}

VVK acknowledges an early conversation with David Berman which has triggered our interest
in instanton contributions to the Lunin-Maldacena duality. 
We thank Chong-Sun Chu and Stefano Kovacs for useful discussions and comments on the manuscript and
 Stefano for helpful remarks on the structure of the relevant correlation functions
in the instanton background.
The research of GG
is supported by PPARC through a Postdoctoral Fellowship. VVK is supported by a PPARC
Senior Fellowship.

\bigskip

\startappendix

\Appendix{D-instanton partition function}

In this Appendix we show construct the D-instanton partition function
in the $\beta$-deformed string theory and show that it reproduces the corresponding
gauge theory result in Section {\bf \ref{sec:insts}}.

In string theory, D-instanton is a point-like defect -- the D(-1) brane,
hence its partition function is described in terms of a matrix model integral.
The partition function of $k$ D-instantons
on $N$ D3-branes in the type IIB theory was previously constructed in Ref.~\cite{MO3}.
Here we will generalize this construction to include the $\beta$-deformation
effects on the string theory side.

Following Ref.~\cite{MO3}
we first consider the $k$D(-1) branes in interacting with the $N$D3 branes
in the standard type IIB string theory. This description accounts for the
D-instanton effects in the undeformed $\AdS_5\times S^5$ background dual to the
$\cN=4$ Yang-Mills.
>From the perspective of the $k$D(-1) world-volume, the
$k$D(-1)/$N$D3 brane system is described by the partition function
\cite{MO3},
\EQ{
\Z_{k,N} = \int d\mu_{k,N} \, e^{-S_{k,N}}\ .
\label{finfin}}
Here the D-instanton integration measure $d\mu_{k,N}$ and action $S_{k,N}$
are over the D-instanton collective coordinates, and the D-3 brane
degrees of freedom are turned off.
This is a $0+0$-dimensional matrix model which can be obtained by the dimensional
reduction
from the $U(k) \times U(N)$
gauge theory describing $k$D$p$ branes and $N$D$(p+4)$ branes.\footnote{From
the instanton perspective, and in the $\alpha' \to 0$ limit, the
overall $U(1)$ factor in the $U(N)$ gauge group is irrelevant.
Hence for the purposes of this paper we will not distinguish between
the $U(k) \times U(N)$ and $U(k) \times SU(N)$ cases. However, the $U(1)$
factor in the $U(k)$ groups is physically significant, it describes
the centre of mass degrees of freedom of the $k$-instanton which are important.}
The $k$D$p$/$N$D$(p+4)$ brane system can live in the maximal dimension
$p=5$ which corresponds to the 6-dimensional gauge theory on the world-volume
of the D5-branes. Then the cases $5\ge p \ge -1$ follow by dimensional reduction.
The D-instanton partition function corresponds to the minimal case of $p=-1$.
For practical calculations it is most convenient to start with the maximal
case $p=5$ to specify the field content of the model, and then reduce to
zero dimensions, $p=-1$.

The content of the $k$D5/$N$D9 system is described by the
$(1,1)$ vector multiplet and two bi-fundamental hypermultiplets in the
6-dimensional world-volume of $k$D5 branes. The vector multiplet transforms in
the adjoint representation of the $U(k)$ gauge group and represents the
open-string degrees of freedom
of the $k$D5 branes in isolation. On the other hand, the $U(k) \times U(N)$
bi-fundamental hyper-multiplets
incorporate the modes of the open strings stretched between the $k$ branes and
the $N$ branes (two species of hypermultiplets correspond to two orientations of the open strings).
Thus the hypermultiplets describe the interactions between D-instantons
and the spectator branes.
The component fields of the vector multiplet are listed in the Table 1, and the
hypermultiplet fields are listed in Table 2.
\begin{table}\setlength{\extrarowheight}{5pt}
\begin{center}\begin{tabular}{||l|l|l|r||} \hline\hline
\phantom{$\Biggr($}
Component & Description & $U(k)$ & $U(N)$ \\
\hline\hline
$\quad\chi^{1 \ldots 6}$ & Gauge Field & ${\Bk}\times{\Bk}$
& ${\bf 1}\quad$
\\
$\quad\lambda_{\aD}$ & Gaugino & ${\Bk}\times{\Bk}$
& ${\bf 1}\quad$
\\
$\quad D^{1 \ldots 3}$ & Auxiliary Field & ${\Bk}\times{\Bk}$
& ${\bf 1}\quad$
\\
\hline
$\quad a^{\prime}_{\alpha \aD}$ & Scalar Field & ${\Bk}\times{\Bk}$
& ${\bf 1}\quad$
\\
$\quad {\cal M}^{\prime}_{\aD}$ & Fermion Field & ${\Bk}\times{\Bk}$
& ${\bf 1}\quad$
\\
\hline\hline
\end{tabular}\end{center}
\caption{\small Components of the $(1,1)$ vector multiplet in $d=6$. They describe $k$ D$5$ branes in
isolation.}
\end{table}

\begin{table}\setlength{\extrarowheight}{5pt}
\begin{center}\begin{tabular}{||l|l|l|r||} \hline\hline
\phantom{$\Biggr($}
Component & Description & $U(k)$ & $U(N)$ \\
\hline\hline
$\quad w_{\aD}$ & Scalar Field & ${\Bk}$ & ${\BN}\quad$
\\
$\quad\mu$ & Fermion Field & ${\Bk}$ & ${\BN}\quad$
\\
\hline
$\quad\bar{w}_{\aD}$ & Scalar Field & ${\bar{\Bk}}$
& ${\bar{\BN}}\quad$
\\
$\quad\bar\mu$ & Fermion Field & ${\bar{\Bk}}$ & ${\bar{\BN}}\quad$
\\
\hline\hline
\end{tabular}\end{center}
\caption{\small Components of bi-fundamental hypermultiplets in $d=6$. They describe interactions
between $k$ D$5$ and $N$ D$9$ branes.}
\end{table}

The D-instanton integration measure is uniquely determined by the action of this $U(k)$ theory
with hypermultiplets. Dimensionally reducing from $d=6$ to $0$ dimensions
one finds  \cite{MO3} the partition function:
\EQ{
\Z_{k,N} = \frac{g_4^4}{{\rm Vol}\,U(k)}
 \int d^{4k^2}a'\,
d^{8k^2}{\cal M}'\, d^{6k^2}\chi\, d^{8k^2}\lambda\, d^{3k^2}D\,
d^{2kN}w\,d^{2kN}\bar w\, d^{4kN}\mu\, d^{4kN}\bar{\mu}\
\exp [-S_{k,N}]
\label{final1}}
where $S_{k,N}=g_0^{-2}S_{G} + S_{K}+S_{D}$ and
\begin{subequations}
\begin{align}
S_{G} & = {\rm tr}_{k}\big(-[\chi_a,\chi_b]^2+\sqrt{2}i\pi
\lambda_{\dot{\alpha}A}[\chi_{AB}^\dagger,\lambda_B^{\dot{\alpha}}]
+2D^{c}D^{c}\big)\, ,\label{p=-1actiona} \\
S_{K} & =  -{\rm tr}_{k}\big([\chi_a,a'_{n}]^2
+\chi_a\bar{w}^\aD_{u}
w_{u\dot{\alpha}}\chi_a + \sqrt{2}i\pi
{\cal M}^{\prime \alpha A}[\chi_{AB},
{\cal M}^{\prime B}_{\alpha}]+2\sqrt{2} i \pi
\bar{\mu}_{u}^{A}\chi_{AB}\mu^{B}_{u}\big)\ ,\label{p=-1actionb} \\
S_{D} & =  i  \pi{\rm tr}_k\big(
[a'_{\alpha\dot{\alpha}},{\cal M}^{\prime\alpha A}]\lambda^{\dot{\alpha}}_{A}
+\bar{\mu}^{A}_{u}w_{u\dot{\alpha}}\lambda^{\dot{\alpha}}_{A}+
\bar{w}_{u\dot{\alpha}}\mu^{A}_{u}
\lambda^{\dot{\alpha}}_{A} + \pi^{-1}D^{c}(\tau^c)^\bD_{\ \aD}
(\bar w^\aD w_\bD+\bar a^{\prime\aD\alpha}a'_{\alpha\bD})\big)\ .
\label{p=-1actionc}
\end{align}\end{subequations}

Equations above define the $k$D-instanton measure in string theory in the flat background and in presence
of the $N$D-3 branes. We now want to $\beta$-deform this background.
Lunin and Maldacena have argued in \cite{LM} that the
open string field theory in the
$\beta$-deformed background is obtained from the theory on the undeformed background
precisely by changing the star-product between the fields carrying the relevant $U(1)$ charges.
In our case, this implies that the star product should be used instead of ordinary products
for all fields transforming under the $SO(6) = U(4)$ R-symmetry. This requires star products
in expressions involving $\chi$, $\lambda$ and $D$ fields in the equations
\eqref{p=-1actiona}-\eqref{p=-1actionc} above. We also recall that the star product is trivial
between the fields of opposite charges and hence can be dropped in the terms which are quadratic
in charged fields. This amounts to the following equations for the action terms in \eqref{final1}
in the $\beta$-deformed background:
\begin{subequations}
\begin{align}
S_{G}^\beta & = {\rm tr}_{k}\big(-[\chi_a,\chi_b]_* [\chi_a,\chi_b]_*
+\sqrt{2}i\pi
\lambda_{\dot{\alpha}A}*[\chi_{AB}^\dagger,\lambda_B^{\dot{\alpha}}]_*
+2D^{c}D^{c}\big)\, ,\label{p=-1Aactiona} \\
S_{K}^\beta & =  -{\rm tr}_{k}\big([\chi_a,a'_{n}]^2
+\chi_a\bar{w}^\aD_{u}
w_{u\dot{\alpha}}\chi_a + \sqrt{2}i\pi
{\cal M}^{\prime \alpha A}*[\chi_{AB},
{\cal M}^{\prime B}_{\alpha}]_*+2\sqrt{2} i \pi
\bar{\mu}_{u}^{A}*\chi_{AB}*\mu^{B}_{u}\big)\ ,\label{p=-1Aactionb} \\
S_{D}^\beta & =  i  \pi{\rm tr}_k\big(
[a'_{\alpha\dot{\alpha}},{\cal M}^{\prime\alpha A}]\lambda^{\dot{\alpha}}_{A}
+\bar{\mu}^{A}_{u}w_{u\dot{\alpha}}\lambda^{\dot{\alpha}}_{A}+
\bar{w}_{u\dot{\alpha}}\mu^{A}_{u}
\lambda^{\dot{\alpha}}_{A} + \pi^{-1}D^{c}(\tau^c)^\bD_{\ \aD}
(\bar w^\aD w_\bD+\bar a^{\prime\aD\alpha}a'_{\alpha\bD})\big)=\, S_{D}
\label{p=-1Aactionc}
\end{align}\end{subequations}

The D-instanton partition function $\Z_{k,N}$
depends explicitly on the inverse string tension $\alpha'$
through the
zero-dimensional coupling $g^{2}_{0}\propto (\alpha')^{-2}$ which appears
in $g_0^{-2}S_{G}^\beta$ which comes from the dimensional reduction of the $d=6$
gauge action.
In the field theory limit the fundamental string scale is set
to zero, $\alpha'=0$, to decouple the world-volume
gauge theory from gravity. Thus, as explained in \cite{MO3},
to derive the ADHM-instanton measure
in conventional supersymmetric gauge theory
one must take the limit $\alpha'\rightarrow 0$.
In this limit $g^{2}_{0}\rightarrow \infty$ equations of motion
for $D^c$ are precisely the non-linear ADHM constraints,
first equation in \eqref{adhmcons}.
Similarly equations of motion for $\lambda$ are the fermionic ADHM
constraints in \eqref{adhmcons}. Integration over $D^c$ and $\lambda^\aD_A$ yields
$\delta$-functions which impose the constraints.

We can now make contact with our result \eqref{partf} for the instanton partition function
in gauge theory derived in
Section {\bf \ref{sec:insts}}. First, we integrate out the
$D^c$ and $\lambda^\aD_A$ variables, thus getting the
$\delta$-functions of the ADHM constraints, precisely as in \eqref{partf}.
Second, we rewrite $S_{K}^\beta$ as,
\begin{equation}
S_{K}^\beta=\, {\rm tr}_k\,\chi_a{\bf L}\chi_a -\, 4\pi i\,
{\rm tr}_{k}\,\chi_{AB}\,\Lambda_{A*B}\ .
\end{equation}
This is equal to (minus) the exponent appearing in equation
\eqref{E53}. On integrating out the gauge field
$\chi_a$, the instanton action reduces to the fermion quadrilinear term
\eqref{E48}. We have therefore reproduced our result for the ADHM measure
in the $\beta$-deformed gauge theory,
up to an overall normalization constant. This is completely analogous
to the matching between D- and gauge-instanton partition functions discovered
in \cite{MO3} -- the only novelty in the present case is the appearance of the
star products on both sides of the correspondence. What this matching really tests in the
$\beta$-deformed theory is the validity of the prescription for introducing $\beta$-deformations
in the open-string theory
conjectured by Lunin and Maldacena and which we have used to derive the results
\eqref{p=-1Aactiona}-\eqref{p=-1Aactionc} above.


\end{document}